\documentclass[11pt]{article}
\usepackage{amssymb,amsmath,amsfonts}
\usepackage{graphicx}
\usepackage{graphics}
\usepackage{eepic,epsfig}

\makeatletter \@addtoreset{equation}{section}

\makeatother

\begin{document}

\title{Vacuum currents induced by a magnetic flux \\
around a cosmic string with finite core}
\author{E. R. Bezerra de Mello$^{1}$\thanks{%
E-mail: emello@fisica.ufpb.br},\, V. B. Bezerra$^{1}$\thanks{%
E-mail: valdir@fisica.ufpb.br}, \, A. A. Saharian$^{2,1}$\thanks{%
E-mail: saharian@ysu.am}, \, H. H. Harutyunyan$^{3}$ \\
\\
\textit{$^{1}$Departamento de F\'{\i}sica, Universidade Federal da Para\'{\i}%
ba}\\
\textit{58.059-970, Caixa Postal 5.008, Jo\~{a}o Pessoa, PB, Brazil}\vspace{%
0.3cm}\\
\textit{$^2$Department of Physics, Yerevan State University,}\\
\textit{1 Alex Manoogian Street, 0025 Yerevan, Armenia}\vspace{0.3cm}\\
\textit{$^3$Department of Mathematics, Armenian State Pedagogical University,%
}\\
\textit{13 Khandjyan Street, 0010 Yerevan, Armenia}}
\maketitle

\begin{abstract}
We evaluate the Hadamard function and the vacuum expectation value of the
current density for a massive complex scalar field in the generalized
geometry of a straight cosmic string with a finite core enclosing an
arbitrary distributed magnetic flux along the string axis. For the interior
geometry, a general cylindrically symmetric static metric tensor is used
with finite support. In the region outside the core, both the Hadamard
function and the current density are decomposed into the idealized
zero-thickness cosmic string and core-induced contributions. The only
nonzero component corresponds to the azimuthal current. The zero-thickness
part of the latter is a periodic function of the magnetic flux inside the
core, with the period equal to the quantum flux. As a consequence of the
direct interaction of the quantum field with the magnetic field inside the
penetrable core, the core-induced contribution, in general, is not a
periodic function of the flux. In addition, the vacuum current, in general,
is not a monotonic function of the distance from the string and may change
the sign. For a general model of the core interior, we also evaluate the
magnetic fields generated by the vacuum current. As applications of the
general results, we have considered an impenetrable core modeled by Robin
boundary condition, a core with the Minkowski-like interior and a core with
a constant positive curvature space. Various exactly solvable distributions
of the magnetic flux are discussed.
\end{abstract}

\bigskip


\bigskip

\section{Introduction}

During the cosmological expansion, the spontaneous breaking of fundamental
symmetries in the early universe leads to phase transitions. In most
interesting models of high energy physics the formation of a variety of
topological defects is predicted as a result of these phase transitions \cite%
{Vile94}. These topologically stable structures have a number of interesting
observable consequences, the detection of which would provide an important
link between cosmology and particle physics. Among the various types of
topological defects the cosmic strings are most thoroughly studied due to
the importance they may have in cosmology. The early interest to this class
of defects was motivated by the scenario in which the strings seed the
primordial density perturbations for the formation of the large-scale
structures in the universe. Although the recent observations of the cosmic
microwave background radiation (CMB) disfavored this scenario, the cosmic
strings are still candidates for the generation of a number interesting
physical effects including the generation of gravitational waves, high
energy cosmic rays, and gamma ray bursts. Among the other signatures are the
gravitational lensing, the creation of small non-Gaussianities in the CMB
and some influence on the corresponding tensor-modes. Recently the cosmic
strings attract a renewed interest partly because a variant of their
formation mechanism is proposed in the framework of brane inflation \cite%
{Sara02}. Note that the cosmic string type conical defects appear also in a
number of condensed matter systems such as crystals, liquid crystals and
quantum liquids (see, for example, Ref. \cite{Nels02}).

In the simplest theoretical model, the spacetime geometry produced by an
infinite straight cosmic string has a conical structure. It is locally flat
except on the top of the string where it has a delta shaped curvature
tensor. In quantum field theory, the non-trivial spatial topology induced by
the presence of a cosmic string raises a number of interesting physical
effects. One of these concerns the effect of a string on the properties of
quantum vacuum. The topology change results in the distortion of the
zero-point vacuum fluctuations of quantized fields and induces shifts in
vacuum expectation values (VEVs) for physical observables. Explicit
calculations have been done for the VEVs of the field squared and
energy-momentum tensor for scalar, fermion and vector fields (see references
given in Ref. \cite{Bell14}). The presence of a magnetic flux along the axis
of the string gives rise to additional polarization of the quantum vacuum
providing another example of a quantum topological interaction \cite{Dowk87}-%
\cite{Site12}. Though the gauge field strength vanishes outside the string
core, the nonvanishing vector potential leads to Aharonov-Bohm-like effects
on the physical characteristics of the vacuum for charged fields. In
particular, in Ref. \cite{Alfo89} it was demonstrated that the generalized
Aharonov-Bohm effect leads to scattering cross sections and to particle
production rates that do not go to zero when the geometrical size of the
string tends to zero. These type of effects induced by gauge field fluxes
have been further investigated in Refs. \cite{Jone10}.

The magnetic flux along the cosmic string induces also vacuum current
densities. This phenomenon has been investigated for scalar fields in Ref.
\cite{Srir01,Site09}. The analysis of induced fermionic currents in
higher-dimensional cosmic string spacetime in the presence of a magnetic
flux have been developed in Ref. \cite{Mello10}. In these analysis the
authors have shown that induced vacuum current densities along the azimuthal
direction appear if the ratio of the magnetic flux by the quantum one has a
nonzero fractional part. Moreover, the fermionic current induced by a
magnetic flux in a $(2+1)$-dimensional conical spacetime and in the presence
of a circular boundary has also been analyzed in Ref. \cite{Saha10}. The VEV
of the current density in the geometry of a cosmic string compactified along
its axis is investigated in Refs. \cite{Beze13} and \cite{Brag14} for
fermionic and scalar fields respectively. The generalization of the
corresponding results for the fermionic case to a cosmic string in
background of de Sitter spacetime is given in Ref. \cite{Moha14}.

Many of treatments of quantum fields around a cosmic string deal mainly with
the case of the idealized geometry where the string is assumed to have zero
thickness. Realistic cosmic strings have internal structure, characterized
by the core radius determined by the symmetry breaking scale at which it is
formed. Cylindrically symmetric static models of a string where the
space-time curvature is spread over a region of nonzero size are considered
in Ref. \cite{Hisk85,Alle90}. The vacuum polarization effects due to
massless fields in the region outside the core are investigated. In
particular, it has been shown that long-range effects can take place due to
the non-trivial core structure. As a model of an impenetrable core for a
cosmic string, in Ref. \cite{Beze06} a cylindrical boundary is considered
with Robin boundary condition on the field operator. The renormalized VEVs
of the field squared and the energy-momentum tensor for a massive scalar
field with general curvature coupling are investigated. The generalization
of the results to the exterior region is given for a general cylindrically
symmetric static model of the string core with finite support.

The present paper is devoted to the investigation of the VEV of the current
density for a massive charged scalar field in the geometry of a cosmic
string with a general cylindrically symmetric core of a finite support. The
core encloses a gauge field flux directed along the string axis with an
arbitrary radial distribution. The current density is among the most
important quantities characterizing the properties od the quantum vacuum.
Though the corresponding operator is local, due to the global nature of the
vacuum, the corresponding VEV describes the global properties of the bulk
and carry information about the structure of the string core. In addition,
the VEV of the current density acts as a source in semiclassical Maxwell
equations and, hence, it plays an important role in modeling a
self-consistent dynamics involving the electromagnetic field. The results
presented below specify the conditions under which the details of the
interior structure can be ignored and the effects induced by cosmic strings
can be approximated by the idealized model.

The organization of the paper is as follows. In the next section we describe
the geometry at hand and present a complete set of mode functions obeying
the matching conditions on the core boundary. By using these mode functions,
in Section \ref{sec:Hadamard}, the Hadamard function is evaluated for the
general case of the core geometry. This function is decomposed into two
parts corresponding to an idealized geometry of a zero thickness cosmic
string and to the finite core contribution. The VEV of the current density
in the region outside the string core is investigated in Section \ref%
{sec:Current}. The behavior of the core-induced contribution is discussed in
various asymptotic regions of the parameters and the corresponding magnetic
fields are evaluated. The Section \ref{sec:Examples} is devoted to
applications of the general results to special cases of the core interior
structure and the magnetic flux distribution. Numerical examples are
presented. The main results of the paper are summarized in Section \ref%
{sec:Conc}. In Appendix \ref{sec:Appendix} we provide an integral
representation for series involving the modified Bessel function. This
representation is used in the evaluation of the Hadamard function and the
current density.

\section{Geometry of the problem and the mode functions}

\label{sec:Geom}

In this paper we consider a $(D+1)$-dimensional background geometry
corresponding to a generalized straight cosmic string with a core of a
finite support having radius $a$. Outside the core, in the cylindrical
coordinate system $(x^{1},x^{2},\ldots ,x^{D})=(r,\phi ,\mathbf{z})$, with $%
\mathbf{z}=(z_{1},\ldots ,z_{N})$ and $N=D-2$, the geometry is given by the
line element with planar angle deficit $2\pi -\phi _{0}$:%
\begin{equation}
ds^{2}=g_{ik}dx^{i}dx^{k}=dt^{2}-dr^{2}-r^{2}d\phi ^{2}-d\mathbf{z}{}^{2},
\label{ds21}
\end{equation}%
where $r>a$, $0\leqslant \phi \leqslant \phi _{0}$, $-\infty <z_{i}<+\infty $
and the points $(r,\phi ,\mathbf{z})$ and $(r,\phi +\phi _{0},\mathbf{z})$
are to be identified. Inside the core, $r<a$, the spacetime geometry is
described by the line element%
\begin{equation}
ds^{2}=dt^{2}-dr^{2}-u^{2}(r)d\phi ^{2}-d\mathbf{z}{}^{2}.  \label{ds2f}
\end{equation}%
By a transformation of the radial coordinate, the line element (\ref{ds2f})
is expressed in the form previously discussed in Ref. \cite{Alle90}. The
value of the radial coordinate corresponding to the symmetry axis of the
core we shall denote by $r=r_{c}$ (as it will be seen below, in general, $%
r_{c}\neq 0$). If the interior geometry is regular on the axis then one has%
\begin{equation}
u(r)\sim q(r-r_{c}),\;q=2\pi /\phi _{0}.  \label{Origin}
\end{equation}%
From the continuity of the metric tensor on the core boundary we get%
\begin{equation}
u(a)=a.  \label{ua}
\end{equation}%
In general, $u^{\prime }(r)$ is not continuous at $r=a$. In the interior
region, $r<a$, for the Ricci scalar one has $R=-2u^{\prime \prime }(r)/u(r)$%
, and in the exterior region $R=0$.

We assume that an additional infinitely thin cylindrical shell located at $%
r=a$ is present with the surface energy-momentum tensor $\tau _{ik}$. Let us
denote by $n^{i}$ the normal to the shell normalized by the condition $%
n_{i}n^{i}=-1$, assuming that it points into the bulk on both sides.
Introducing the induced metric as $h_{ik}=g_{ik}+n_{i}n_{k}$, for the
extrinsic curvature tensor of the separating boundary, we have $%
K_{ik}=h_{i}^{l}h_{k}^{p}\nabla _{l}n_{p}$. The latter is related to the
surface energy-momentum tensor by the Israel matching condition%
\begin{equation}
\left\{ K_{ik}-Kh_{ik}\right\} =8\pi G\tau _{ik},  \label{matchcond}
\end{equation}%
where $K=K_{i}^{i}$ and the curly brackets denote summation over each side
of the shell. In the problem under consideration, for the interior and
exterior regions one has $n_{i}=\delta _{i}^{1}$ and $n_{i}=-\delta _{i}^{1}$%
, respectively. For the only nonzero component of the extrinsic curvature
tensor we get $K_{2}^{2}=-u_{a}^{\prime }/a$ and $K_{2}^{2}=1/a$, with $%
u_{a}^{\prime }=u^{\prime }(a)$, for the interior and exterior regions,
respectively. Now, the condition (\ref{matchcond}) leads to the following
surface energy-momentum tensor:
\begin{equation}
8\pi G\tau _{i}^{k}=\frac{u_{a}^{\prime }-1}{a}\delta
_{i}^{k},\;i,k=0,3,\ldots ,D,  \label{Surf}
\end{equation}%
and $\tau _{1}^{1}=\tau _{2}^{2}=0$.

In this paper we are interested in the VEV of the current density for a
charged scalar field $\varphi (x)$ in the region outside the string core.
For the field with curvature coupling parameter $\xi $, the field equation
has the form
\begin{equation}
(g^{ik}D_{i}D_{k}+m^{2}+\xi R)\varphi (x)=0,  \label{fieldeq}
\end{equation}%
where $D_{k}=\nabla _{k}+ieA_{k}$, with $e$ being the charge associated with
the field, and $\nabla _{i}$ the covariant derivative operator. The values
of the curvature coupling parameter $\xi =0$ and $\xi =\xi _{D}\equiv
(D-1)/4D$ correspond to the most important special cases of minimally and
conformally coupled scalars, respectively. The operator of the current
density is given by the expression
\begin{equation}
j_{l}(x)=ie[\varphi ^{+}(x)D_{l}\varphi (x)-(D_{l}\varphi (x))^{+}\varphi
(x)].  \label{jl}
\end{equation}%
The corresponding VEV can be expressed in terms of the Hadamard function%
\begin{equation}
G(x,x^{\prime })=\left\langle 0\right\vert \varphi (x)\varphi ^{+}(x^{\prime
})+\varphi ^{+}(x^{\prime })\varphi (x)\left\vert 0\right\rangle ,
\label{G1}
\end{equation}%
where $\left\vert 0\right\rangle $ stands for the vacuum state. For the VEV
one has
\begin{equation}
\left\langle j_{l}(x)\right\rangle =\frac{i}{2}e\lim_{x^{\prime }\rightarrow
x}(\partial _{l}-\partial _{l}^{\prime }+2ieA_{l})G(x,x^{\prime }).
\label{jlVEV}
\end{equation}

For the evaluation of the Hadamard function we shall use the mode summation
formula%
\begin{equation}
G(x,x^{\prime })=\sum_{\alpha }\sum_{s=\pm }\varphi _{\alpha
}^{(s)}(x)\varphi _{\alpha }^{(s)\ast }(x^{\prime }),  \label{G11}
\end{equation}%
where $\varphi _{\alpha }^{(\pm )}(x)$ is a complete set of normalized
positive- and negative-energy mode functions obeying the field equation (\ref%
{fieldeq}) and specified by the collective index $\alpha $. First we
consider the mode functions for scattering states. In accordance with the
symmetry of the problem, these functions can be expressed as
\begin{equation}
\varphi _{\alpha }^{(\pm )}(x)=f(r)e^{iqn\phi +i\mathbf{kz}\mp i\omega t},
\label{phialf}
\end{equation}%
where $n=0,\pm 1,\pm 2,\cdots $, $\mathbf{k}=(k_{1},\ldots ,k_{N})$, $%
-\infty <k_{j}<\infty $. Substituting into the field equation, and assuming
the vector potential of the form
\begin{equation}
A_{l}=(0,0,A_{2}(r),0,\ldots ,0),  \label{Ak}
\end{equation}%
for the radial function in the interior region we obtain the following
equation
\begin{equation}
\left\{ u(r)\partial _{1}[u(r)\partial _{1}]+u^{2}(r)\gamma
^{2}-[qn+eA_{2}(r)]^{2}-\xi u^{2}(r)R\right\} f(r)=0.  \label{radeq}
\end{equation}%
where%
\begin{equation}
\gamma =\sqrt{\omega ^{2}-k^{2}-m^{2}}.  \label{gan}
\end{equation}%
For the interior region $\;R=-2u^{\prime \prime }/u$. In the exterior
region, $r>a$, the corresponding equation is obtained taking in Eq. (\ref%
{radeq}) $u(r)=r$ and $R=0$. Note that in Eq. (\ref{radeq}), $A_{2}(r)$ is
the covariant component of the vector potential. For the physical component
one has $A_{\phi }(r)=-A_{2}(r)/u(r)$ for $r<a$ and $A_{\phi
}(r)=-A_{2}(r)/r $ for $r>a$. The only nonzero components of the field
tensor $F_{il}$ corresponding to Eq. (\ref{Ak}) are given by $%
F_{12}=-F_{21}=\partial _{r}A_{2}(r)$. In $D=3$ models this corresponds to a
magnetic field directed along the $z$-axis with the strength $%
B_{z}=-r^{-1}\partial _{r}A_{2}(r)$.

The radial functions in the interior and exterior regions are connected by
the matching conditions at the boundary. The radial function is continuous
at the boundary: $f(a-)=f(a+)$. In order to find the jump condition for its
first derivative, we note that the Ricci scalar contains a delta function
term $2(u_{a}^{\prime }-1)\delta (r-a)/a$ located on the separating
boundary. By taking into account this term and integrating the radial
equation near the boundary we find the following condition
\begin{equation}
f^{\prime }(a+)-f^{\prime }(a-)=2\xi (u_{a}^{\prime }-1)f(a)/a.  \label{jump}
\end{equation}%
In deriving this condition we have assumed that the vector potential
contains no Dirac delta-type terms.

In what follows we shall consider the case when the magnetic field in the
exterior region vanishes, corresponding to $A_{2}(r)=A_{2}=\mathrm{const}$
for $r>a$. In this case, for the exterior region, the general solution of
the radial equation is a linear combination of the Bessel and Neumann
functions $J_{\beta _{n}}(\gamma r)$ and $Y_{\beta _{n}}(\gamma r)$ of the
order
\begin{equation}
\beta _{n}=q|n+\beta |,\;\beta =eA_{2}/q.  \label{betn}
\end{equation}%
Note that, if the function $A_{2}(r)$ is regular, the parameter $\beta $ is
expressed in terms of the magnetic flux inside the core as $\beta =-\Phi
/\Phi _{0}$, with $\Phi _{0}=2\pi /e$ being the quantum flux. Let us denote
by $R_{n}(r,\gamma )$ the solution of the radial equation (\ref{radeq})
inside the core, regular on the axis, namely at $r=r_{c}$. By taking into
account that the parameter $\gamma $ enters in the equation in the form $%
\gamma ^{2}$, without loss of generality we can take this function to be
real and obeying the condition $R_{n}(r,\gamma e^{\pi i})=\mathrm{const\,}%
R_{n}(r,\gamma )$. By the same reason, if $A_{2}(r)=0$ for $r<a$, we can
also assume that $R_{-n}(r,\gamma )=\mathrm{const\,}R_{n}(r,\gamma )$.

Now, the radial part of the eigenfunctions can be written in the form%
\begin{equation}
f(r)=\left\{
\begin{array}{ll}
R(r,\gamma ), & r<a, \\
AJ_{\beta _{n}}(\gamma r)+BY_{\beta _{n}}(\gamma r), & r>a.%
\end{array}%
\right.  \label{fr}
\end{equation}%
The coefficients $A$ and $B$ in Eq. (\ref{fr}) are determined from the
condition of the continuity of the radial functions and from the jump
condition (\ref{jump}) at $r=a$:%
\begin{eqnarray}
A &=&\frac{\pi }{2}R_{n}(a,\gamma )Y_{\beta _{n}}(\gamma a,p_{n}(\gamma )),
\notag \\
B &=&-\frac{\pi }{2}R_{n}(a,\gamma )J_{\beta _{n}}(\gamma a,p_{n}(\gamma )),
\label{AB}
\end{eqnarray}%
where
\begin{equation}
p_{n}(\gamma )=a\frac{R_{n}^{\prime }(a,\gamma )}{R_{n}(a,\gamma )}+2\xi
(u_{a}^{\prime }-1).  \label{P}
\end{equation}%
with $R_{n}^{\prime }(a,\gamma )=\partial _{r}R_{n}(r,\gamma )|_{r=a}$. Here
and in what follows, for a given function $F(x)$, we use the notation%
\begin{equation}
F(x,y)=xF^{\prime }(x)-yF(x).  \label{Fxy}
\end{equation}%
Note that, because of our choice of the function $R_{n}(r,\gamma )$, we have
the property%
\begin{equation}
p_{n}(\gamma e^{\pi i})=p_{n}(\gamma ).  \label{pnProp}
\end{equation}%
In addition, if $A_{2}(r)=0$ for $r<a$, one has $p_{-n}(\gamma
)=p_{n}(\gamma )$.

From the normalization condition for the mode functions one has%
\begin{equation}
\int d^{D}x\,u(r)\varphi _{\alpha }^{(\lambda )}(x)\varphi _{\alpha ^{\prime
}}^{(\lambda ^{\prime })\ast }(x)=\frac{1}{2\omega }\delta _{\alpha \alpha
^{\prime }}\delta _{\lambda \lambda ^{\prime }},  \label{Norm}
\end{equation}%
where $u(r)=r$ in the region $r>a$, the symbol $\delta _{\alpha \alpha
^{\prime }}$ is understood as Kronecker delta for discrete indices and as
the Dirac delta function for continuous ones. For the radial functions we
get
\begin{equation}
(2\pi )^{N}\phi _{0}\int_{r_{c}}^{\infty }dr\,u(r)f(r,\gamma )f^{\ast
}(r,\gamma ^{\prime })=\frac{1}{2\omega }\delta (\gamma -\gamma ^{\prime }),
\label{NormRad}
\end{equation}%
where $f(r,\gamma )$ is defined by the right-hand side of Eq. (\ref{fr}).
The integral over the interior region is finite and, hence, the dominant
contribution in Eq. (\ref{NormRad}) for $\gamma =\gamma ^{\prime }$ comes
from the integration over the exterior region. By using the standard
integrals involving the cylinder functions, we find the relation
\begin{equation}
(\pi /2)^{2}R^{2}(a,\gamma )=\frac{(2\pi )^{-N}\gamma }{2\phi _{0}\omega }%
\left[ J_{\beta _{n}}^{2}(\gamma a,p_{n}(\gamma ))+Y_{\beta _{n}}^{2}(\gamma
a,p_{n}(\gamma ))\right] ^{-1}.  \label{NormR}
\end{equation}%
This relation determines the normalization of the interior radial function.
With this normalization, the radial function in the exterior region, $r>a$,
is expressed as%
\begin{equation}
f(r)=\left[ \frac{(2\pi )^{-N}\gamma }{2\phi _{0}\omega }\right] ^{1/2}\frac{%
g_{\beta _{n}}(\gamma r,\gamma a,p_{n}(\gamma ))}{[J_{\beta _{n}}^{2}(\gamma
a,p_{n}(\gamma ))+Y_{\beta _{n}}^{2}(\gamma a,p_{n}(\gamma ))]^{1/2}},
\label{fext}
\end{equation}%
where we use the notation%
\begin{equation}
g_{\beta _{n}}(z,x,y)=Y_{\beta _{n}}(x,y)J_{\beta _{n}}(z)-J_{\beta
_{n}}(x,y)Y_{\beta _{n}}(z).  \label{gbet}
\end{equation}

In addition to the scattering states one can have bound states for which $%
\gamma $ is purely imaginary, $\gamma =i\chi $, $\chi >0$. In order to have
a stable vacuum we assume that $\chi <m$, otherwise the modes with imaginary
energy would be present. For the bound states the exterior radial function
is expressed in terms of the MacDonald function:%
\begin{equation}
f(r)=A_{b}K_{\beta _{n}}(\chi r),\;r>a.  \label{fb}
\end{equation}%
From the continuity condition for the radial function at $r=a$, we find%
\begin{equation}
A_{b}=R_{n}(a,i\chi )/K_{\beta _{n}}(\chi a).  \label{Ab}
\end{equation}%
The jump condition for the radial derivative gives%
\begin{equation}
K_{\beta _{n}}(\chi a,p_{n}(i\chi ))=0,  \label{bseq}
\end{equation}%
with the notation defined by Eq. (\ref{Fxy}). The solutions of Eq. (\ref%
{bseq}) for $\chi $ determine the possible bound states.

The normalization condition for the radial functions of the bound states is
given by Eq. (\ref{NormRad}), where in the right-hand side we should take
the Kronecker delta $\delta _{\chi \chi ^{\prime }}$ and the energy is given
by $\omega =\sqrt{k^{2}+m^{2}-\chi ^{2}}$. Now, both the interior and
exterior regions contribute to the normalization integral and one needs the
formula for the integrals involving the function $R_{n}(r,\gamma )$. From
Eq. (\ref{radeq}) it follows that for the two solutions $f_{i}(r,\gamma )$ ($%
i=1,2$) of that equation, one has%
\begin{equation}
\int dr\,u(r)f_{1}(r,\gamma )f_{2}(r,\gamma ^{\prime })=\frac{u(r)}{\gamma
^{2}-\gamma ^{\prime 2}}\left[ f_{1}(r,\gamma )\partial _{r}f_{2}(r,\gamma
^{\prime })-f_{2}(r,\gamma ^{\prime })\partial _{r}f_{1}(r,\gamma )\right] .
\label{IntForm1}
\end{equation}%
From here we get%
\begin{equation}
\int_{r_{c}}^{a}dr\,u(r)R_{n}^{2}(r,\gamma )=\frac{a}{2\gamma }[(\partial
_{\gamma }R_{n}(a,\gamma ))R_{n}^{\prime }(a,\gamma )-R_{n}(a,\gamma
)\partial _{\gamma }R_{n}^{\prime }(a,\gamma )].  \label{IntForm2}
\end{equation}%
For the integral over the exterior region we use the standard integral
involving the square of the MacDonald function \cite{Prud86}. By taking into
account Eq. (\ref{bseq}) for the bound states, the normalization integral
can be presented as%
\begin{equation}
\int_{r_{c}}^{a}dr\,u(r)R_{n}^{2}(r,i\chi )+A_{b}^{2}\int_{a}^{\infty
}dr\,rK_{\beta _{n}}^{2}(\chi r)=-\frac{A_{b}^{2}}{2\chi }K_{\beta
_{n}}(\chi a)\partial _{\chi }K_{\beta _{n}}(\chi a,p_{n}(i\chi )).
\label{NormInt}
\end{equation}%
Now, from the normalization condition one gets%
\begin{equation}
R_{n}^{2}(a,i\chi )=-\frac{q\chi K_{\beta _{n}}(\chi a)}{(2\pi )^{D-1}\omega
\partial _{\chi }K_{\beta _{n}}(\chi a,p_{n}(i\chi ))}.  \label{Normbs}
\end{equation}%
The coefficient $A_{b}$ in Eq. (\ref{fb}) is found from Eq. (\ref{Ab}).

\section{Hadamard function}

\label{sec:Hadamard}

After the construction of the mode functions we turn to the evaluation of
the Hadamard function in the exterior region. First we consider the case
when the bound states are absent. Plugging the functions (\ref{phialf}) with
(\ref{fext}) into the mode-sum (\ref{G11}), we find the following expression
for the Hadamard function
\begin{eqnarray}
G(x,x^{\prime }) &=&\frac{1}{\phi _{0}}\int \frac{d^{N}{\mathbf{k}}}{(2\pi
)^{N}}\sum_{n=-\infty }^{+\infty }e^{i{\mathbf{k}}\Delta {\mathbf{z+}}%
iqn\Delta \phi }\int_{0}^{\infty }d\gamma \,\frac{\cos (\omega \Delta t)}{%
\omega }  \notag \\
&&\times \frac{\gamma g_{\beta _{n}}(\gamma r,\gamma a,p_{n}(\gamma
))g_{\beta _{n}}(\gamma r^{\prime },\gamma a,p_{n}(\gamma ))}{J_{\beta
_{n}}^{2}(\gamma a,p_{n}(\gamma ))+Y_{\beta _{n}}^{2}(\gamma a,p_{n}(\gamma
))},  \label{G12}
\end{eqnarray}%
where $\Delta {\mathbf{z=z}}-{\mathbf{z}}^{\prime }$, $\Delta \phi =\phi
-\phi ^{\prime }$, $\Delta t=t-t^{\prime }$ and $\omega =\sqrt{\gamma
^{2}+k^{2}+m^{2}}$. For the further transformation of this formula we use
the identity
\begin{equation}
\frac{g_{\beta _{n}}(\gamma r,\gamma a,p_{n}(\gamma ))g_{\beta _{n}}(\gamma
r^{\prime },\gamma a,p_{n}(\gamma ))}{J_{\beta _{n}}^{2}(\gamma
a,p_{n}(\gamma ))+Y_{\beta _{n}}^{2}(\gamma a,p_{n}(\gamma ))}=J_{\beta
_{n}}(\gamma r)J_{\beta _{n}}(\gamma r^{\prime })-\frac{1}{2}\sum_{s=1}^{2}%
\frac{J_{\beta _{n}}(\gamma a,p_{n}(\gamma ))}{H_{\beta _{n}}^{(s)}(\gamma
a,p_{n}(\gamma ))}H_{\beta _{n}}^{(s)}(\gamma r)H_{\beta _{n}}^{(s)}(\gamma
r^{\prime }),  \label{Ident}
\end{equation}%
where $H_{\beta _{n}}^{(s)}(z)$, $s=1,2$, are the Hankel functions and,
again, we use the notation (\ref{Fxy}). Substituting Eq. (\ref{Ident}) into
Eq. (\ref{G12}), in the part with the second term in the right-hand side of
Eq. (\ref{Ident}), in the complex plane $\gamma $, we rotate the integration
contour by the angle $\pi /2$ for $s=1$ and by the angle $-\pi /2$ for $s=2$%
. By using the property (\ref{pnProp}), one can see that the integrals over
the segments $(0,i\sqrt{k^{2}+m^{2}})$ and $(0,-i\sqrt{k^{2}+m^{2}})$ are
canceled. In the remaining integral over the imaginary axis we introduce the
modified Bessel functions. As a result, the Hadamard function is presented
in the decomposed form%
\begin{equation}
G(x,x^{\prime })=G_{0}(x,x^{\prime })+G_{c}(x,x^{\prime }),  \label{Gdec}
\end{equation}%
where%
\begin{eqnarray}
G_{0}(x,x^{\prime }) &=&\frac{1}{\phi _{0}}\int \frac{d^{N}{\mathbf{k}}}{%
(2\pi )^{N}}\sum_{n=-\infty }^{+\infty }e^{i{\mathbf{k}}\Delta {\mathbf{z+}}%
iqn\Delta \phi }\int_{0}^{\infty }d\gamma   \notag \\
&&\times \frac{\gamma }{\omega }J_{\beta _{n}}(\gamma r)J_{\beta
_{n}}(\gamma r^{\prime })\cos (\omega \Delta t),  \label{G0}
\end{eqnarray}%
and%
\begin{eqnarray}
G_{c}(x,x^{\prime }) &=&-\frac{2}{\pi \phi _{0}}\int \frac{d^{N}\mathbf{k}}{%
(2\pi )^{N}}\sum_{n=-\infty }^{+\infty }e^{i\mathbf{k}\Delta \mathbf{z+}%
iqn\Delta \phi }\int_{0}^{\infty }dx  \notag \\
&&\times \cosh (x\Delta t)K_{\beta _{n}}(zr)K_{\beta _{n}}(zr^{\prime })%
\frac{I_{\beta _{n}}(za,p_{n}(iz))}{K_{\beta _{n}}(za,p_{n}(iz))},
\label{Gc}
\end{eqnarray}%
with $z=\sqrt{x^{2}+k^{2}+m^{2}}$. In Eq. (\ref{Gc}), we have used the
notation (\ref{Fxy}) with $F(x)=I_{\beta _{n}}(x)$ being the modified Bessel
function.

\begin{figure}[tbph]
\begin{center}
\epsfig{figure=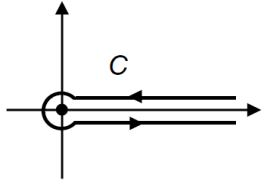,width=3.5cm,height=2.5cm}
\end{center}
\caption{The contour of the integration in the representation (\protect\ref%
{IntRepCos}).}
\label{fig1}
\end{figure}

The equation (\ref{G0}) corresponds to the Hadamard function in the geometry
of an idealized cosmic string with zero-thickness core and constant vector
potential with the regularity condition imposed at the origin. For the
transformation of this function we use the integral representation \cite%
{Beze14}%
\begin{equation}
\frac{\cos (\omega \Delta t)}{\omega }=-\frac{1}{2\sqrt{\pi }}\int_{C}\frac{%
ds}{s^{1/2}}\,e^{-\omega ^{2}s+(\Delta t)^{2}/(4s)},  \label{IntRepCos}
\end{equation}%
with the contour of the integration depicted in figure \ref{fig1}. The
integral can also be presented in the form $\int_{C}ds=\int_{c_{\rho
}}ds-2\int_{\rho }^{\infty }ds$, where $c_{\rho }$ is a circle of radius $%
\rho $ with the center at the origin of the complex $s$-plane and having the
anticlockwise direction.

After the substitution of Eq. (\ref{IntRepCos}) into Eq. (\ref{G0}), the
integrals are evaluated by using the formulas%
\begin{equation}
\int \frac{d^{N}{\mathbf{k}}}{(2\pi )^{N}}\,e^{i\mathbf{k}\cdot \Delta
\mathbf{z}-sk^{2}}=\frac{e^{-|\Delta \mathbf{z}|^{2}/(4s)}}{(4\pi
)^{N/2}s^{N/2}},  \label{Int1}
\end{equation}%
and \cite{Prud86}%
\begin{equation}
\int_{0}^{\infty }d\gamma \gamma J_{\beta _{n}}(\gamma r)J_{\beta
_{n}}(\gamma r^{\prime })\,e^{-s\gamma ^{2}}=\frac{1}{2s}\exp \left( -\frac{%
r^{2}+r^{\prime 2}}{4s}\right) I_{\beta _{n}}(rr^{\prime }/2s).  \label{Int2}
\end{equation}%
In the remaining integral over the contour $C$, under the condition $|\Delta
\mathbf{z}|^{2}+r^{2}+r^{\prime 2}>(\Delta t)^{2}$, the part corresponding
to the integral over the circle $c_{\rho }$ vanishes in the limit $\rho
\rightarrow 0$. Thus, taking this limit one gets%
\begin{equation}
G_{0}(x,x^{\prime })=q\frac{(rr^{\prime })^{-(D-1)/2}}{(2\pi )^{(D+1)/2}}%
\int_{0}^{\infty }dx\,x^{(D-3)/2}e^{-m^{2}rr^{\prime }/(2x)-x\left(
r^{2}+r^{\prime 2}+|\Delta \mathbf{z}|^{2}-(\Delta t)^{2}\right)
/(2rr^{\prime })}\mathcal{I}_{q}(\beta ,\Delta \phi ,x),  \label{G0b}
\end{equation}%
where we have introduced a new integration variable $x=rr^{\prime }/s$ and
the function%
\begin{equation}
\mathcal{I}_{q}(\beta ,\Delta \phi ,x)=\sum_{n=-\infty }^{+\infty
}e^{iqn\Delta \phi }I_{\beta _{n}}(x).  \label{Ical}
\end{equation}

An alternative expression for the Hadamard function $G_{0}(x,x^{\prime })$
is obtained by using the integral representation (\ref{Ical1}) for the
function (\ref{Ical}). After the integration over $x$ we get%
\begin{eqnarray}
&&G_{0}(x,x^{\prime })=\frac{2m^{D-1}}{(2\pi )^{(D+1)/2}}\left\{
\sum_{n}e^{2\pi in\beta }f_{(D-1)/2}(ms_{n}(x,x^{\prime }))-\frac{%
qe^{-iqn_{0}\Delta \phi }}{2\pi i}\sum_{j=\pm 1}je^{ji\pi q\beta _{f}}\right.
\notag \\
&&\qquad \times \left. \int_{0}^{\infty }dy\,\frac{\cosh [qy(1-\beta
_{f})]-\cosh (q\beta _{f}y)e^{-iq\left( \Delta \phi +j\pi \right) }}{\cosh
(qy)-\cos [q\left( \Delta \phi +j\pi \right) ]}f_{(D-1)/2}(mc(y,x,x^{\prime
}))\right\} ,  \label{G0c}
\end{eqnarray}%
where the summation over $n$ goes under the same conditions as in the
right-hand side of Eq. (\ref{1term}),
\begin{eqnarray}
s_{n}(x,x^{\prime }) &=&[(\Delta r)^{2}+|\Delta \mathbf{z}|^{2}-(\Delta
t)^{2}+4rr^{\prime }\sin ^{2}\left( \pi n/q-\Delta \phi /2\right) ]^{1/2},
\notag \\
c(y,x,x^{\prime }) &=&\left[ (\Delta r)^{2}+|\Delta \mathbf{z}|^{2}-(\Delta
t)^{2}+4rr^{\prime }\cosh ^{2}(y/2)\right] ^{1/2},  \label{sc}
\end{eqnarray}%
and we have defined the function%
\begin{equation}
f_{\nu }(x)=K_{\nu }(x)/x^{\nu }.  \label{fnu}
\end{equation}%
Note that the $n=0$ term in Eq. (\ref{G0c}) corresponds to the Hadamard
function in Minkowski spacetime in the absence of the magnetic flux. The
divergences in the coincidence limit of the arguments are contained in this
term only. The remaining part is induced by the planar angle deficit and by
the magnetic flux. Note that the Euclidean Green function for a massive
scalar field in the geometry of a zero thickness cosmic string has been
considered in Ref. \cite{Guim94} in the special case $q<2$ and in the
subspace $-1+q/2<\Delta \phi /\phi _{0}<1-q/2$. Under these conditions, the
only contribution in the series over $n$ in Eq. (\ref{G0c}) comes from the
term $n=0$.

Now, let us discuss the contribution to the Hadamard function coming from
possible bound states. The latter are solutions of Eq. (\ref{bseq}) and the
corresponding mode functions are given by Eq. (\ref{fb}). The contribution
of the bound states to the Hadamard function is given by
\begin{eqnarray}
G_{b}(x,x^{\prime }) &=&-\frac{2}{\phi _{0}}\int \frac{d^{N}{\mathbf{k}}}{%
(2\pi )^{N}}\sum_{n=-\infty }^{+\infty }e^{i{\mathbf{k}}\Delta {\mathbf{z+}}%
iqn\Delta \phi }\sum_{\chi }\,\frac{\chi }{\omega }  \notag \\
&&\times \frac{\cos (\omega \Delta t)K_{\beta _{n}}(\eta r)K_{\beta
_{n}}(\eta r^{\prime })}{K_{\beta _{n}}(\chi a)\partial _{\chi }K_{\beta
_{n}}(\chi a,p_{n}(i\chi ))},  \label{Gb}
\end{eqnarray}%
where $\sum_{\chi }$ stands for the summation over the solutions of Eq. (\ref%
{bseq}) and $\omega =\sqrt{k^{2}+m^{2}-\chi ^{2}}$. In the presence of the
bound states, the evaluation of the part in the Hadamard function
corresponding to scattering modes is similar to what we have described
before. The only difference appears in the part where we rotate the
integration contour in the contribution coming from the second term in the
right-hand side of Eq. (\ref{Ident}). Now, the integrand has simple poles $%
\gamma =\pm i\chi $ on the imaginary axis corresponding to the zeros of the
function $H_{\beta _{n}}^{(s)}(\gamma a,p_{n}(\gamma ))$. We escape these
poles by semicircles of small radius in the right half-plane. The
contributions from the semicircles in the upper and lower half-planes
combine into the residue of the term with $s=1$ at $i\chi $ multiplied by $%
2\pi i$. As a result, an additional term comes from the poles, given by%
\begin{equation}
G_{\mathrm{poles}}(x,x^{\prime })=\frac{2}{\phi _{0}}\int \frac{d^{N}{%
\mathbf{k}}}{(2\pi )^{N}}\sum_{n=-\infty }^{+\infty }e^{i{\mathbf{k}}\Delta {%
\mathbf{z+}}iqn\Delta \phi }\sum_{\chi }\frac{\chi }{\omega }\frac{\cos
(\omega \Delta t)I_{\beta _{n}}(\chi a,p_{n}(i\chi ))}{\partial _{\chi
}K_{\beta _{n}}(\chi a,p_{n}(i\chi ))}K_{\beta _{n}}(\chi r)K_{\beta
_{n}}(\chi r^{\prime }).  \label{Gpol}
\end{equation}%
From the Wronskian relation for the modified Bessel functions, by using Eq. (%
\ref{bseq}), one can show that%
\begin{equation}
I_{\beta _{n}}(\chi a,p(i\chi ))=1/K_{\beta _{n}}(\chi a).  \label{RelIK}
\end{equation}%
With this relation, we see that the contribution coming from the poles on
the imaginary axis cancel the part (\ref{Gb}) corresponding to the bound
states. As a result, the expression for the Hadamard function (\ref{Gc}) is
valid in the presence of bound states as well.

\section{Current density}

\label{sec:Current}

With a given Hadamard function, we can evaluate the VEV of the current
density by using Eq. (\ref{jlVEV}). The current density is decomposed as
\begin{equation}
\left\langle j_{l}\right\rangle =\left\langle j_{l}\right\rangle
_{0}+\left\langle j_{l}\right\rangle _{c},  \label{jldec}
\end{equation}%
where the part $\left\langle j_{l}\right\rangle _{0}$ is the VEV in the
geometry of zero thickness cosmic string and the contribution $\left\langle
j_{l}\right\rangle _{c}$ is induced by the nontrivial structure of the core.
The only nonzero component corresponds to the azimuthal current with $l=2$.

First we consider the current density for the geometry of an idealized
cosmic string. By using the Hadamard function given by Eq. (\ref{G0b}), for
the corresponding physical component $\left\langle j_{\phi }\right\rangle
_{0}=-\left\langle j_{2}\right\rangle _{0}/r$, one finds%
\begin{equation}
\left\langle j_{\phi }\right\rangle _{0}=\frac{eq^{2}r^{-D}}{(2\pi
)^{(D+1)/2}}\int_{0}^{\infty
}dx\,x^{(D-3)/2}e^{-x-m^{2}r^{2}/(2x)}\sum_{n=-\infty }^{+\infty }(n+\beta
)I_{\beta _{n}}(x).  \label{j20}
\end{equation}%
As it is seen, in the geometry of a zero thickness cosmic string the current
density is a periodic function of the parameter $\beta $ with period 1. This
corresponds to the periodicity of the current as a function of the magnetic
flux with the period equal to the quantum flux. So, if we present the
parameter $\beta $ in the form%
\begin{equation}
\beta =n_{0}+\beta _{f},\;0\leqslant \beta _{f}<1,  \label{betdec}
\end{equation}%
with $n_{0}$ being an integer, then the VEV, $\left\langle j_{\phi
}\right\rangle _{0}$, is a function of $\beta _{f}$ alone. For $\beta _{f}=0$
this VEV vanishes.

An equivalent expression for the current density is obtained with the help
of the integral representation (\ref{Isum2}), given in the appendix:
\begin{eqnarray}
\left\langle j_{\phi }\right\rangle _{0} &=&\frac{4em^{D+1}r}{(2\pi
)^{(D+1)/2}}\left\{ \sum_{n=1}^{[q/2]}\sin (2\pi n\beta )\sin \left( 2\pi
n/q\right) f_{(D+1)/2}(2mr\sin (\pi n/q))\right.  \notag \\
&&\left. +\frac{q}{2\pi }\int_{0}^{\infty }dy\,f_{(D+1)/2}(2mr\cosh (y/2))%
\frac{\sinh yf(q,\beta _{f},y)}{\cosh (qy)-\cos (\pi q)}\right\} ,
\label{j20b}
\end{eqnarray}%
where the function $f(q,\beta ,y)$ is defined by the relation
\begin{equation}
f(q,\beta _{f},y)=\sin (\pi q\beta _{f})\sinh [q(1-\beta _{f})y]-\sin [\pi
q(1-\beta _{f})]\sinh (q\beta _{f}y).  \label{fq}
\end{equation}%
In the special case $1\leqslant q\leqslant 2$ the sum over $n$ is absent and
Eq. (\ref{j20b}) is reduced to the expression given in Ref. \cite{Site09}
(see also \cite{Brag14}). Note that the current density for an idealized
geometry of zero thickness cosmic string does not depend on the curvature
coupling parameter. At large distances from the string, $mr\gg 1$, and for $%
q>2$ the dominant contribution comes from the term with $n=1$ and the
current density decays as $e^{-2mr\sin (\pi /q)}/(mr)^{D/2}$. For $%
q\leqslant 2$, the suppression of the VEV at large distances is stronger, by
the factor $e^{-2mr}/(mr)^{(D+3)/2}$.

For a massless field Eq. (\ref{j20b}) is reduced to
\begin{eqnarray}
\left\langle j_{\phi }\right\rangle _{0} &=&\frac{4e\Gamma ((D+1)/2)}{(4\pi
)^{(D+1)/2}r^{D}}\left\{ \sum_{n=1}^{[q/2]}\frac{\sin (2\pi n\beta )\cos
\left( \pi n/q\right) }{\sin ^{D}(\pi n/q)}\right.   \notag \\
&&\left. +\frac{q}{\pi }\int_{0}^{\infty }dy\,\frac{\sinh y}{\cosh ^{D}y}%
\frac{f(q,\beta _{f},2y)}{\cosh (2qy)-\cos (\pi q)}\right\} .  \label{j20m0}
\end{eqnarray}%
For a massive field, the expression in the right-hand side of Eq. (\ref%
{j20m0}) gives the leading term in the asymptotic expansion of $\left\langle
j_{\phi }\right\rangle _{0}$ for points close to the string, $mr\ll 1$. In
particular, the current density diverges on the string as $1/r^{D}$. Another
special case corresponds to the absence of the planar angle deficit, $q=1$.
From Eq. (\ref{j20b}), we get%
\begin{equation}
\left\langle j_{\phi }\right\rangle _{0}=\frac{16em^{D+1}r}{(2\pi )^{(D+3)/2}%
}\sin (\pi \beta _{f})\int_{0}^{\infty }dy\,\sinh y\sinh [(1-2\beta
_{f})y]f_{(D+1)/2}(2mr\cosh y).  \label{j20q1}
\end{equation}%
In this case the azimuthal current density is positive for $0<\beta _{f}<1/2$
and negative for $1/2<\beta _{f}<1$. After the change of the integration
variable, it can be seen that Eq. (\ref{j20q1}) coincides with the result
obtained in Ref. \cite{Site98}.

\begin{figure}[tbph]
\begin{center}
\epsfig{figure=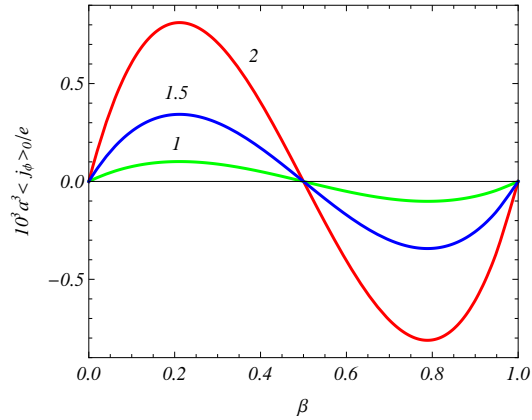,width=7.cm,height=5.5cm}
\end{center}
\caption{The azimuthal current density in the geometry of a zero thickness
cosmic string, as a function of the parameter $\protect\beta $, for a $D=3$
massless scalar field. The graphs are plotted for $r/a=2$ and the numbers
near the curves correspond to the values of $q$.}
\label{fig2}
\end{figure}

In figure \ref{fig2}, the VEV of the current density, measured in units of $%
a $, $a^{3}\left\langle j_{\phi }\right\rangle _{0}$, is displayed as a
function of the parameter $\beta $ for a massless scalar field in a
3-dimensional space ($D=3$). The graphs are plotted for a fixed value of the
distance from the string, corresponding to $r/a=2$, and the numbers near the
curves correspond to the values of the parameter $q$.

For the core-induced contribution in Eq. (\ref{jldec}), by using Eq. (\ref%
{Gc}) for the corresponding part of the Hadamard function, we get
\begin{equation}
\left\langle j_{\phi }\right\rangle _{c}=-8eq^{2}\frac{(4\pi )^{-(D+1)/2}}{%
\Gamma ((D-1)/2)r}\sum_{n=-\infty }^{+\infty }(n+\beta )\int_{m}^{\infty
}dx\,x(x^{2}-m^{2})^{(D-3)/2}\frac{I_{\beta _{n}}(ax,p_{n}(ix))}{K_{\beta
_{n}}(ax,p_{n}(ix))}K_{\beta _{n}}^{2}(rx).  \label{j2c}
\end{equation}%
If $A_{2}(r)=0$ for $r<a$, one has $p_{-n}(\gamma )=p_{n}(\gamma )$ and this
contribution vanishes for $\beta _{f}=0$. Note that the function $p_{n}(ix)$%
, in general, depends on the parameter $\beta $ through the boundary
condition for the vector potential at the core boundary. As a consequence of
this, the core-induced contribution to the current density, in general, is
not a periodic function of the magnetic flux enclosed by the core. The
physical reason for this is the interaction of the quantum field with the
magnetic field inside the core if the latter is penetrable. Substituting Eq.
(\ref{betdec}) into Eq. (\ref{j2c}) and redefining the summation variable,
the expression is written in an alternative form
\begin{eqnarray}
\left\langle j_{\phi }\right\rangle _{c} &=&-8eq^{2}\frac{(4\pi )^{-(D+1)/2}%
}{\Gamma ((D-1)/2)r}\sum_{n=-\infty }^{+\infty }(n+\beta
_{f})\int_{m}^{\infty }dx\,x  \notag \\
&&\times (x^{2}-m^{2})^{(D-3)/2}\frac{I_{\beta _{fn}}(ax,p_{n-n_{0}}(ix))}{%
K_{\beta _{fn}}(ax,p_{n-n_{0}}(ix))}K_{\beta _{fn}}^{2}(rx),  \label{j2cb}
\end{eqnarray}%
where $\beta _{fn}=q|n+\beta _{f}|$. The only dependence on $n_{0}$ appears
in the function $p_{n-n_{0}}(ix)$. Note the the dependence of the function $%
p_{n-n_{0}}(ix)$ on $n_{0}$ can also come through the parameter $\beta $.
For $\beta _{f}=0$ the part $\left\langle j_{\phi }\right\rangle _{0}$
vanishes and the appearance of the nonzero current density is a purely
finite-core effect.

Let us consider the behavior of the core-induced part at large distances
from the core. For a massive field and for $mr\gg 1$, the dominant
contribution into Eq. (\ref{j2c}) comes from the region near the lower limit
of integration. To the leading order we get
\begin{equation}
\left\langle j_{\phi }\right\rangle _{c}\approx -\frac{eq^{2}m^{D}}{2(4\pi
)^{(D-1)/2}}\frac{e^{-2rm}}{(mr)^{(D+3)/2}}\sum_{n=-\infty }^{+\infty
}(n+\beta )\frac{I_{\beta _{n}}(am,p_{n}(im))}{K_{\beta _{n}}(am,p_{n}(im))}.
\label{j2cmfar}
\end{equation}%
and the core-induced part is suppressed exponentially. Comparing with the
behavior of the part $\left\langle j_{\phi }\right\rangle _{0}$, we see that
for $q>2$ the latter dominates at large distances and the relative
contribution of the finite core-induced effects are suppressed by the factor
$e^{-2mr[1-\sin (\pi /q)]}/(mr)^{3/2}$. For $q\leqslant 2$, the
contributions of $\left\langle j_{\phi }\right\rangle _{0}$ and $%
\left\langle j_{\phi }\right\rangle _{c}$ to the total VEV, at large
distances, are of the same order if $am\sim 1$.

For a massless field and at large distances from the core boundary,
introducing in Eq. (\ref{j2cb}) a new integration variable, $y=rx$, we
expand the integrand in powers of $a/r$. For $\beta _{f}\neq 0$, the
dominant contribution comes from the term with $n=n_{\beta }$, where $%
n_{\beta }=0$ for $0<\beta _{f}<1/2$ and $n_{\beta }=-1$ for $1/2<\beta
_{f}<1$. To the leading order we get%
\begin{equation}
\left\langle j_{\phi }\right\rangle _{c}\approx \mathrm{sgn}(1/2-\beta _{f})%
\frac{2eq\Gamma (2\sigma _{\beta }+(D-1)/2)}{(4\pi
)^{D/2}r^{D}(2r/a)^{2\sigma _{\beta }}}\frac{\sigma _{\beta }-p_{n_{\beta
}-n_{0}}(0)}{\sigma _{\beta }+p_{n_{\beta }-n_{0}}(0)}\frac{\Gamma (\sigma
_{\beta }+(D-1)/2)}{\Gamma ^{2}(\sigma _{\beta })\Gamma (\sigma _{\beta
}+D/2)},  \label{j2cm0far}
\end{equation}%
with the notation
\begin{equation}
\sigma _{\beta }=\left\{
\begin{array}{ll}
q\beta _{f}, & 0<\beta _{f}<1/2, \\
q(1-\beta _{f}), & 1/2<\beta _{f}<1.%
\end{array}%
\right.  \label{sigma}
\end{equation}%
In this case the relative contribution of the core-induced effects are
suppressed by the factor $(a/r)^{2\sigma _{\beta }}$. For $\beta _{f}=0$ the
dominant contribution comes from the terms with $n=\pm 1$:%
\begin{equation}
\left\langle j_{\phi }\right\rangle _{c}\approx \frac{2eq\Gamma (2q+(D-1)/2)%
}{(4\pi )^{D/2}r^{D}(2r/a)^{2q}}\sum_{n=\pm 1}n\frac{q-p_{n-n_{0}}(0)}{%
q+p_{n-n_{0}}(0)}\frac{\Gamma (q+(D-1)/2)}{\Gamma ^{2}(q)\Gamma (q+D/2)}.
\label{j2cm0bet0far}
\end{equation}%
In this case $\left\langle j_{\phi }\right\rangle _{0}=0$ and the only
effect comes from the nontrivial core structure.

Now let us consider the behavior of the core-induced part in the current
density near the core boundary. At the boundary, in general, the current
density diverges. For points near the boundary the dominant contribution in
Eq. (\ref{j2c}) comes from large values of $|n|$. In order to find the
leading term in the asymptotic expansion over the distance from the core
boundary, it is convenient to introduce in Eq. (\ref{j2c}) a new integration
variable $y=x/\beta _{n}$. In what follows we shall need the uniform
asymptotic expansion of the function $p_{n}(i\beta _{n}y)$ for large values
of $|n|$. First we consider the corresponding expansion for the function
\begin{equation}
y_{n}(r,x)=\frac{R_{n}^{\prime }(r,ix)}{R_{n}(r,ix)}.  \label{yn}
\end{equation}%
By using Eq. (\ref{radeq}) for the interior radial function, the following
equation is obtained for the function (\ref{yn}):%
\begin{equation}
\frac{1}{u(r)}[u(r)y_{n}(r,x)]^{\prime }+y_{n}^{2}(r,x)-x^{2}-\frac{%
[qn+eA_{2}(r)]^{2}}{u^{2}(r)}+2\xi \frac{u^{\prime \prime }(r)}{u(r)}=0,
\label{Eqyn}
\end{equation}%
where the prime stands for the derivative over $r$. Substituting $x=\beta
_{n}y$, to the leading order over $|n|$, one finds from Eq. (\ref{Eqyn}) the
result
\begin{equation}
y_{n}(r,\beta _{n}y)=\pm \beta _{n}\sqrt{y^{2}+1/u^{2}(r)}.  \label{yn1}
\end{equation}%
For the regular solution the upper sign should be taken. Putting Eq. (\ref%
{yn1}) in the first term of Eq. (\ref{Eqyn}) (with $x=\beta _{n}y$), we can
find the next to the leading term which is given by%
\begin{equation}
ay_{n}(a,\beta _{n}y)\approx \beta _{n}\sqrt{1+a^{2}y^{2}}\left[ 1+\frac{%
U(ay)}{\beta _{n}}+\cdots \right] ,  \label{yn2}
\end{equation}%
where
\begin{equation}
U(x)=\mathrm{sgn}(n)\frac{eA_{2a}-q\beta }{x^{2}+1}-\frac{x^{2}u_{a}^{\prime
}}{2(x^{2}+1)^{3/2}},  \label{Ux}
\end{equation}%
and $A_{2a}=\lim_{r\rightarrow a-}A_{2}(r)$. The next terms in the expansion
over the powers of $1/\beta _{n}$ are found in a similar way. The
corresponding expansion for the function $p_{n}(i\beta _{n}y)$ is obtained
from Eq. (\ref{P}):
\begin{equation}
p_{n}(i\beta _{n}y)\approx \sqrt{1+a^{2}y^{2}}\left[ \beta _{n}+U(ay)\right]
+2\xi (u_{a}^{\prime }-1).  \label{pn2}
\end{equation}

Now, by using the uniform asymptotic expansions for the modified Bessel
functions for large values of the order \cite{Abra72}, we can see that, to
the leading order,
\begin{equation}
\frac{I_{\beta _{n}}(\beta _{n}ay,p(i\beta _{n}y))}{K_{\beta _{n}}(\beta
_{n}ay,p(i\beta _{n}y))}K_{\beta _{n}}^{2}(\beta _{n}ry)\approx \frac{%
e^{-2\beta _{n}\sqrt{a^{2}y^{2}+1}(r/a-1)}}{4\beta _{n}^{2}(a^{2}y^{2}+1)}%
V(ay),  \label{Asymp}
\end{equation}%
where%
\begin{equation}
V(x)=\left( 2\xi -\frac{x^{2}/2}{x^{2}+1}\right) (u_{a}^{\prime }-1)+\mathrm{%
sgn}(n)\frac{eA_{2a}-q\beta }{\sqrt{x^{2}+1}}.  \label{Vx}
\end{equation}%
Note that the leading term in the expansion of the function $I_{\beta
_{n}}(\beta _{n}ay,p(i\beta _{n}y))$ vanishes due to the cancelation of the
leading terms for the functions $\beta _{n}ayI_{\beta _{n}}^{\prime }(\beta
_{n}ay)$ and $p(i\beta _{n}y)I_{\beta _{n}}(\beta _{n}ay)$.

Substituting Eq. (\ref{Asymp}) into Eq. (\ref{j2c}) (with $x$ replaced by $%
\beta _{n}y$), we first evaluate the series over $n$ (to the leading order
over $r-a$). Then, the integral over $y$ is expressed in terms of the gamma
function. Two cases should be considered separately. In the case $%
eA_{2a}\neq q\beta $, to the leading order one gets
\begin{equation}
\left\langle j_{\phi }\right\rangle _{c}\approx e\frac{q\beta -eA_{2a}}{%
(4\pi )^{(D+1)/2}}\frac{\Gamma ((D-1)/2)}{Da(r-a)^{D-1}}.  \label{j2cnear}
\end{equation}%
The condition $eA_{2a}\neq q\beta $ means that the vector potential is not
continuous at the core boundary and the corresponding magnetic field
contains the part with the delta type distribution located on the
cylindrical shell $r=a$. For $eA_{2a}=q\beta $ the vector potential is
continuous and the leading term vanishes. The next to the leading term in
the expansion of the core-induced contribution behaves as $(r/a-1)^{2-D}$.
The corresponding coefficient depends on $A_{2}^{\prime }(a)$ and $u^{\prime
\prime }(a)$.

The analysis given above shows that in the models of the core for which the
core-induced contribution in the current density diverges on the boundary,
the latter dominates for points near the boundary, whereas at large
distances the contribution of the part $\left\langle j_{\phi }\right\rangle
_{0}$ dominates. Depending on the core model, these two contributions may
have different signs and, as a result the total VEV is not a monotonic
function of the radial coordinate and changes the sign for some value of the
latter. An example of this type of behavior will be given below in the
flower-pot model for the core geometry.

It is of interest to investigate the asymptotic behavior of the core-induced
part in the current density for large values of the magnetic flux. To this
aim the form of the current density (\ref{j2cb}) is more convenient. From
Eq. (\ref{pn2}) it follows that for a fixed value of $x$ and for large $|n|$
one has $p_{n}(ix)\approx q|n|$. Hence, in models where $p_{n}(ix)$ does not
depend on $\beta $ (for an example see below), for fixed $x$ and $n$, to the
leading order we get $p_{n-n_{0}}(ix)\approx q|n_{0}|$. By taking into
account that the dominant contribution to the series and integral in Eq. (%
\ref{j2cb}) comes from the regions $|n|\lesssim 1/(r/a-1)$ and $x\lesssim
1/(r-a)$, we see that the latter estimate is valid for $|n_{0}|\gg 1/(r/a-1)$%
. Under this condition, in Eq. (\ref{j2cb}) we can use the following
replacement
\begin{equation}
\frac{I_{\beta _{fn}}(ax,p_{n-n_{0}}(ix))}{K_{\beta
_{fn}}(ax,p_{n-n_{0}}(ix))}\approx \frac{I_{\beta _{fn}}(ax)}{K_{\beta
_{fn}}(ax)}.  \label{repl}
\end{equation}%
In this case, for large values of the magnetic flux the core-induced
contribution in the current density coincides, to the leading order, with
the corresponding result for a cylindrical shell with Dirichlet boundary
condition (see below). In particular, it is a periodic function of the
magnetic flux. In models where the function $p_{n}(ix)$ depends on $\beta $,
an additional dependence on $n_{0}$ appears in the function $p_{n-n_{0}}(ix)$%
. As a result of this, a leading term proportional to $|n_{0}|$ may vanish
(an example of this type of situation will be given below). The next to the
leading term does not depend on $n_{0}$ and, again, the corresponding
contribution to the current density is periodic in the magnetic flux.

Now we consider the core-induced contribution in the current density in the
limit $a\rightarrow 0$ for fixed value of $\beta $. The latter means that
the magnetic flux inside the core is fixed. The dominant contribution to Eq.
(\ref{j2cb}) comes from the term with $n=0$ in the case $0<\beta _{f}<1/2$
and from the term $n=-1$ in the case $1/2<\beta _{f}<1$. To the leading
order we find%
\begin{eqnarray}
\left\langle j_{\phi }\right\rangle _{c} &\approx &16\,\mathrm{sgn}%
(1/2-\beta _{f})eq^{2}\frac{(4\pi )^{-(D+1)/2}}{\Gamma ((D-1)/2)r^{D}}\frac{%
(a/2r)^{2\sigma _{\beta }}}{\Gamma ^{2}(\sigma _{\beta })}  \notag \\
&&\times \int_{mr}^{\infty }dx\,x^{2\sigma _{\beta
}+1}(x^{2}-m^{2}r^{2})^{(D-3)/2}\frac{\sigma _{\beta
}-p_{n-n_{0}}^{(0)}(ix/r)}{\sigma _{\beta }+p_{n-n_{0}}^{(0)}(ix/r)}%
K_{\sigma _{\beta }}^{2}(r),  \label{j2a0}
\end{eqnarray}%
where $p_{n-n_{0}}^{(0)}(ix)=\lim_{a\rightarrow 0}p_{n-n_{0}}(ix)$ and $%
\sigma _{\beta }$ is defined by Eq. (\ref{sigma}). In models where the
limiting value $p_{n-n_{0}}^{(0)}(ix)=p_{n-n_{0}}^{(0)}$ does not depend on $%
x$, for a massless field the leading term is given by Eqs. (\ref{j2cm0far})
and (\ref{j2cm0bet0far}). Note that for integer values of $\beta $ the term
with $\beta _{n}=0$ does not contribute to the core-induced contribution in
the current and, hence, the long-range effects of the core, discussed in
Ref. \cite{Alle90} for the VEV of the field squared, are absent for the
current density.

The field strength for the electromagnetic field, $F_{ik}$, generated by the
vacuum current density is found from the semiclassical Maxwell equation $%
\nabla _{k}F^{ik}=-4\pi \left\langle j^{i}\right\rangle $. In the problem
under consideration, the only nonzero components correspond to $%
F^{12}=-F^{21}$. After the simple integration, from the Maxwell equation for
the contravariant component we get%
\begin{equation}
F^{21}=\frac{4\pi }{r}\int_{r}^{\infty }dr\,\left\langle j_{\phi
}\right\rangle .  \label{F21}
\end{equation}%
In the model with $D=3$ the corresponding magnetic field $\mathbf{B}$ is
directed along the string axis and $B_{z}=rF^{21}$. Similar to the current
density, the field strength is decomposed into the zero thickness cosmic
string and core-induced contributions:%
\begin{equation}
F^{21}=F_{(0)}^{21}+F_{(c)}^{21}.  \label{F21dec}
\end{equation}%
By using the integration formula $\int_{r}^{\infty }dr\,rf_{\nu
}(br)=b^{-2}f_{\nu -1}(br)$, for the idealized cosmic string part we get%
\begin{eqnarray}
F_{(0)}^{21} &=&\frac{4em^{D-1}}{(2\pi )^{(D-1)/2}r}\left\{
\sum_{n=1}^{[q/2]}\sin (2\pi n\beta )\cot \left( \pi n/q\right)
f_{(D-1)/2}(2mr\sin (\pi n/q))\right.   \notag \\
&&\left. +\frac{q}{\pi }\int_{0}^{\infty }dy\,\frac{\tanh yf(q,\beta _{f},2y)%
}{\cosh (2qy)-\cos (\pi q)}\,f_{(D-1)/2}(2mr\cosh y)\right\} .  \label{F210}
\end{eqnarray}%
In the case $1\leqslant q<2$, the corresponding magnetic field has been
investigated in Ref. \cite{Site09}.

For the core-induced contribution to the field strength, by using Eq. (\ref%
{j2cb}), one obtains%
\begin{eqnarray}
F_{(c)}^{21} &=&-4eq\frac{(4\pi )^{-(D-1)/2}}{\Gamma ((D-1)/2)}%
\sum_{n=-\infty }^{+\infty }\mathrm{sgn}(n+\beta _{f})\int_{m}^{\infty
}dx\,x^{2}  \notag \\
&&\times (x^{2}-m^{2})^{(D-3)/2}\frac{I_{\beta _{fn}}(ax,p_{n-n_{0}}(ix))}{%
K_{\beta _{fn}}(ax,p_{n-n_{0}}(ix))}G_{\beta _{fn}}(xr),  \label{F21c}
\end{eqnarray}%
where we have introduced the function%
\begin{equation}
G_{\nu }(y)=K_{\nu }^{\prime }(y)\partial _{\nu }K_{\nu }(y)-K_{\nu
}(y)\partial _{\nu }K_{\nu }^{\prime }(y).  \label{Gnu}
\end{equation}%
In deriving Eq. (\ref{F21c}) we have used the formula from Ref. \cite{Prud86}
for an indefinite integral involving the product of the MacDonald functions
and applied the l'H\^{o}pital's rule.

\section{Examples of the core model}

\label{sec:Examples}

In the discussion above we have considered a general model specified by the
functions $u(r)$ and $A_{2}(r)$ in the region $r<a$. In this section we
discuss exactly solvable examples of these functions. Note that, in general,
the radii of the string core and of the gauge field flux can be different.
For example, in the abelian Higgs model these radii are determined by the
Higgs and vector particle masses (see, for instance, Ref. \cite{Vile94}).
Though our analysis described above is valid for general case, in the
examples below for simplicity we assume that the radii of the flux tube and
string core coincide. In particular, the latter is the case for BPS cosmic
strings. The VEVs of the field squared and energy-momentum tensor for a
charged scalar field in the geometry of an idealized cosmic string with zero
thickness core and in the presence of a finite radius magnetic flux have
been considered in Ref. \cite{Spin03}.

\subsection{Cylindrical shell with Robin boundary condition}

First we consider a model with an impenetrable core, on the boundary of
which the field obeys the Robin boundary condition:
\begin{equation}
\left( \partial _{r}-\sigma \right) \varphi =0,\quad r=a,  \label{Rbc}
\end{equation}%
with a constant parameter $\sigma $. In this case the VEVs in the exterior
region do not depend on the interior geometry and are completely determined
by the total flux, enclosed by the boundary, through the parameter $\beta $
and by the Robin coefficient $\sigma $. For a neutral scalar field, the
corresponding VEVs of the field squared and the energy-momentum tensor
inside and outside a cylindrical shell with Robin boundary condition have
been investigated in Ref. \cite{Beze06} (for boundary-induced quantum vacuum
effects in the geometry of a cosmic string see Ref. \cite{Brev95}). For a
charged field and in the presence of a magnetic flux, the mode functions in
the exterior region have the form (\ref{phialf}) with the radial function $%
f(r)$ being the linear combination of the functions $J_{\beta _{n}}(\gamma r)
$ and $Y_{\beta _{n}}(\gamma r)$. The relative coefficient is determined
from the boundary condition (\ref{Rbc}) and for the mode functions one has%
\begin{equation}
\varphi _{\alpha }^{(\pm )}(x)=\beta _{\alpha }g_{\beta _{n}}(\gamma
r,\gamma a,\sigma a)e^{iqn\phi +i\mathbf{kz}\mp i\omega t},  \label{phiRob}
\end{equation}%
where the function $g_{\beta _{n}}(z,x,y)$ is defined in Eq. (\ref{gbet}).

The normalization coefficient is determined from the condition (\ref{Norm})
where now the integration goes over the exterior region. This gives
\begin{equation}
\beta _{\alpha }^{2}=\frac{q(2\pi )^{1-D}\gamma /(2\omega )}{J_{\beta
_{n}}^{2}(\gamma a,\sigma a)+Y_{\beta _{n}}^{2}(\gamma a,\sigma a)}.
\label{betR}
\end{equation}%
We see that for the Robin shell the mode functions in the exterior region
are obtained from the corresponding functions for the general model of the
core taking%
\begin{equation}
p_{n}(\gamma )=\sigma a.  \label{Robrepl}
\end{equation}%
Note that, similar to the case of the zero thickness cosmic string, here the
current density does not depend on the curvature coupling parameter.

The expressions for the Hadamard function and for the VEV of the current
density are obtained from the formulae given above by the substitution (\ref%
{Robrepl}). Now, the function $p_{n}(ix)$ in Eq. (\ref{j2c}) does not depend
on $n$ and $\beta $. From here it follows that the core-induced contribution
and the total VEV of the current density are periodic functions of the
magnetic flux with the period equal to the quantum flux. The equation (\ref%
{bseq}) for the bound states is reduced to $yK_{\beta _{n}}^{\prime
}(y)/K_{\beta _{n}}(y)=\sigma a$ with $y=\chi a$. By taking into account
that $yK_{\beta _{n}}^{\prime }(y)/K_{\beta _{n}}(y)<-\beta _{n}$, we
conclude that there are no bound states for $\sigma a\geqslant -q\,\mathrm{%
min}(\beta _{f},1-\beta _{f})$. For points near the boundary the leading
term in the asymptotic expansion for the current density is obtained in a
way similar to that we have described above in the general case. By using
the uniform asymptotic expansion of the modified Bessel functions \cite%
{Abra72}, we can see that the coefficient of the leading term vanishes.

\begin{figure}[tbph]
\begin{center}
\epsfig{figure=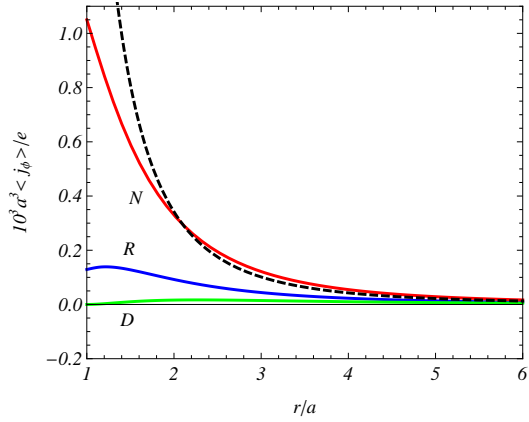,width=7.cm,height=5.5cm}
\end{center}
\caption{The current density for a $D=3$ massless scalar field with
Dirichlet, Neumann and Robin boundary conditions on the core boundary, as a
function of $r/a$, for fixed values $\protect\beta =0.2$ and $q=1.5$. For
Robin boundary condition $\protect\sigma a=1$. The dashed line presents the
current density in the geometry of a zero thickness cosmic string. }
\label{fig3}
\end{figure}

Figure \ref{fig3} displays the dependence of the total current density on
the radial coordinate for $D=3$ massless scalar field with Dirichlet,
Neumann and Robin boundary conditions. For Robin boundary condition we have
taken $\sigma a=1$ and in all cases the graphs are plotted for fixed values $%
\beta =0.2$, $q=1.5$. The dashed line presents the current density for a
zero thickness cosmic string, namely, $10^{3}a^{3}\left\langle j_{\phi
}\right\rangle _{0}/e$. For the corresponding value on the boundary one has $%
\left\langle j_{\phi }\right\rangle _{0}\approx 2.7\cdot 10^{-3}e/a^{3}$.
Numerical results show that the subleading term in the expansion of the
core-induced current density over the distance from the boundary vanishes as
well and the current is finite on the boundary.

\begin{figure}[tbph]
\begin{center}
\begin{tabular}{cc}
\epsfig{figure=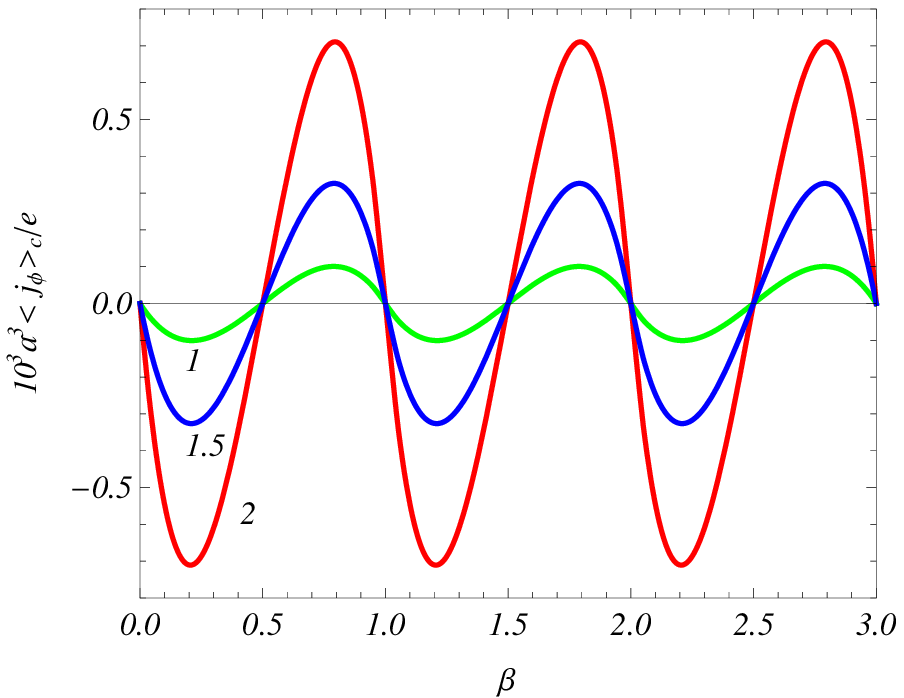,width=7.cm,height=5.5cm} & \quad %
\epsfig{figure=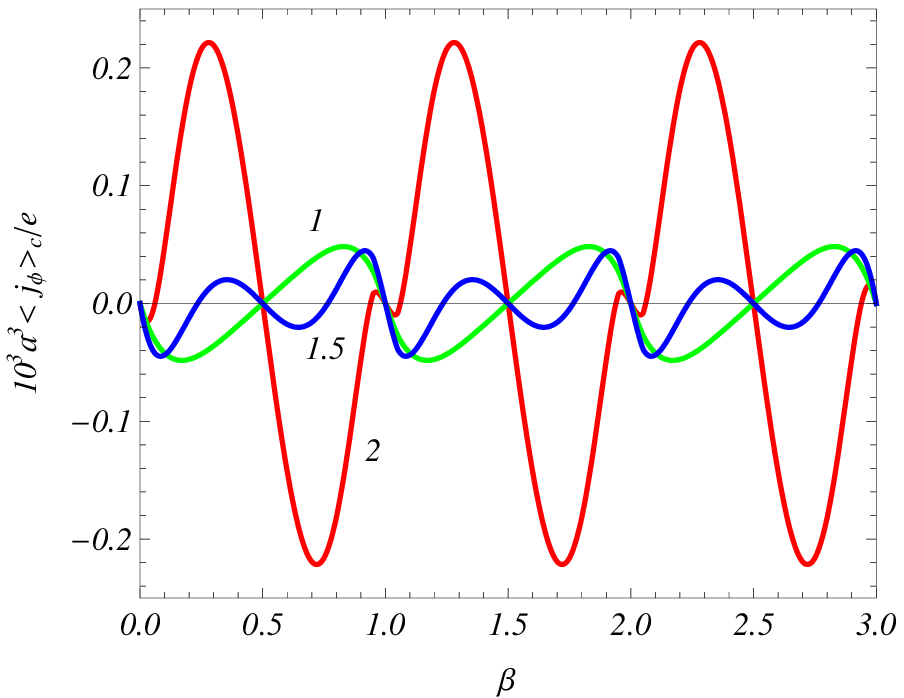,width=7.cm,height=5.5cm}%
\end{tabular}%
\end{center}
\caption{The core-induced contribution in the current density for a $D=3$
massless scalar field versus the parameter $\protect\beta $. The left and
right panels correspond to the model of impenetrable core with Dirichlet and
Neumann boundary conditions, respectively. The numbers near the curves
present the corresponding values of $q$ and the graphs are plotted for $%
r/a=2 $.}
\label{fig4}
\end{figure}

In figure \ref{fig4}, for $D=3$ massless scalar field and for $r/a=2$, we
have plotted the core-induced contribution to the VEV of the current density
as a function of $\beta $ for different values of the parameter $q$ (numbers
near the curves). The left and right panels correspond to Dirichlet and
Neumann boundary conditions on the core boundary.

\subsection{Flower-pot model}

In the flower-pot model, discussed in Ref. \cite{Alle90}, the interior
geometry is Minkowskian. Introducing a new angular coordinate $\phi $, $%
0\leqslant \phi \leqslant 2\pi /q$, related to the Minkowskian one by $\phi
_{M}=q\phi $, the interior line element is written as%
\begin{equation}
ds_{M}^{2}=dt^{2}-dr_{M}^{2}-q^{2}r_{M}^{2}d\phi ^{2}-d\mathbf{z}{}^{2},
\label{FPint}
\end{equation}%
with $0\leqslant r_{M}\leqslant r_{Ma}$, where $r_{Ma}$ is the corresponding
value at the core boundary (see below). From the continuity of the radial
component of the metric tensor at the boundary of the core we have $r_{M}=r+%
\mathrm{const}$. From the continuity of the $g_{22}$ component one gets $%
\mathrm{const}=a(1/q-1)$. Hence, for the function $u(r)$ in Eq. (\ref{ds2f}%
), we obtain%
\begin{equation}
r_{M}=r-r_{c},\;u(r)=q(r-r_{c}),\;r_{c}=a(1-1/q),  \label{FPu}
\end{equation}%
with $r\geqslant r_{c}$. Note that $r_{Ma}=r_{M}|_{r=a}=a/q$, i.e., in terms
of the radial coordinate $r_{M}$ the radius of the core is equal to $a/q$.
For the nonzero components of the surface energy-momentum tensor one has%
\begin{equation}
8\pi G\tau _{i}^{k}=\frac{q-1}{a}\delta _{i}^{k},\;i,k=0,3,\ldots ,D,
\label{FPemt}
\end{equation}%
and the corresponding energy density is positive for $q>1$. We should also
specify the function $A_{2}(r)$ inside the core. Here we consider examples
where the equation for the radial parts of the mode functions is exactly
integrable.

First we consider the gauge field configuration corresponding to Dirac delta
function type field strength located on the boundary of the core. The
Aharonov-Bohm scattering of particles on this type of flux distribution has
been discussed in Refs. \cite{Hage90,Bord93,Andr12}. The corresponding
vector potential is given by the expression
\begin{equation}
A_{2}(r)=A_{2}\theta (r-a).  \label{A2shell}
\end{equation}%
In $D=3$ models this corresponds to the magnetic field strength $%
B_{z}=-A_{2}\delta (r-a)/a$. The regular solution in the interior region is
given by
\begin{equation}
R_{n}(r,\gamma )=CJ_{n}(\gamma r_{M}),  \label{Rshell}
\end{equation}%
and the coefficient $C$ is determined from the relation (\ref{NormR}). For
the function in the expression (\ref{j2c}) of the current density one gets%
\begin{equation}
p_{n}(ix)=ax\frac{I_{n}^{\prime }(xa/q)}{I_{n}(xa/q)}+2\xi (q-1).
\label{pixshell}
\end{equation}%
This function does not depend on the parameter $\beta $. In accordance with
the asymptotic analysis presented in the previous section, for large values
of the magnetic flux, $|n_{0}|\gg 1/(r/a-1)$, the leading term in the
expansion of the core-induced VEV coincides with that for a cylindrical
shell with Dirichlet boundary condition. This term is periodic in the
magnetic flux with the period equal to the quantum flux. From the relation $%
yI_{n}^{\prime }(y)/I_{n}(y)\geqslant |n|$ it follows that the bound states
are absent if the condition $2\xi (1-1/q)\geqslant -\beta _{f}$ is
satisfied. In particular, this is the case for minimally and conformally
coupled fields.

The second example we want to consider corresponds to a homogeneous magnetic
field inside the core with the vector potential%
\begin{equation}
A_{2}(r)=-qA^{(2)}r_{M}^{2}.  \label{A2homo}
\end{equation}%
This model of the magnetic flux tube in the context of the Aharonov-Bohm
effect has been considered in Ref. \cite{Bord93,Sere85}. From the continuity
at the core boundary one gets $A^{(2)}=-qA_{2}/a^{2}$, or in terms of the
parameter $\beta $: $eA^{(2)}=-\beta q^{2}/a^{2}$. In the Minkowskian
coordinates $(t,r_{M},\phi _{M},\mathbf{z})$, the nonzero component of the
vector potential is given by $A_{2}(r)/q$. In $D=3$ model, the strength of
the corresponding magnetic field is expressed as $B_{z}=-2\beta
q^{2}/(ea^{2})$. For the regular solution inside the core one has
\begin{equation}
R_{n}(r,\gamma )=\frac{C_{1}r_{M}^{|n|}}{e^{|\beta |(r_{M}q/a)^{2}/2}}%
M\left( \frac{1+|n|}{2}-\frac{(\gamma a/q)^{2}-2n\beta }{4|\beta |}%
,1+|n|,|\beta |(r_{M}q/a)^{2}\right) ,  \label{Rnhomo}
\end{equation}%
where $M(a,b,x)=\,_{1}F_{1}(a;b;x)$ is the confluent hypergeometric function
\cite{Abra72} and $C_{1}$ is determined from Eq. (\ref{NormR}). For the
function $p(ix)$ in Eq. (\ref{j2c}) this gives%
\begin{eqnarray}
p_{n}(ix) &=&q|n|-q|\beta |+2\xi (q-1)+q|\beta |\frac{1+|n|+2\kappa (x)}{%
1+|n|}  \notag \\
&&\times \frac{M((3+|n|)/2+\kappa (x),2+|n|,|\beta |)}{M((1+|n|)/2+\kappa
(x),1+|n|,|\beta |)},  \label{pnxhom}
\end{eqnarray}%
where we have defined%
\begin{equation}
\kappa (x)=\frac{(ax/q)^{2}+2n\beta }{4|\beta |}.  \label{kapc}
\end{equation}%
For large values of the magnetic flux, $|n_{0}|\gg 1/(r/a-1)$, the leading
term in the expansion of Eq. (\ref{pnxhom}), which is proportional to $%
|n_{0}|$, vanishes. The next to the leading term does not depend on $n_{0}$
and, to the leading order, the current density is a periodic function of the
magnetic flux.

\begin{figure}[tbph]
\begin{center}
\begin{tabular}{cc}
\epsfig{figure=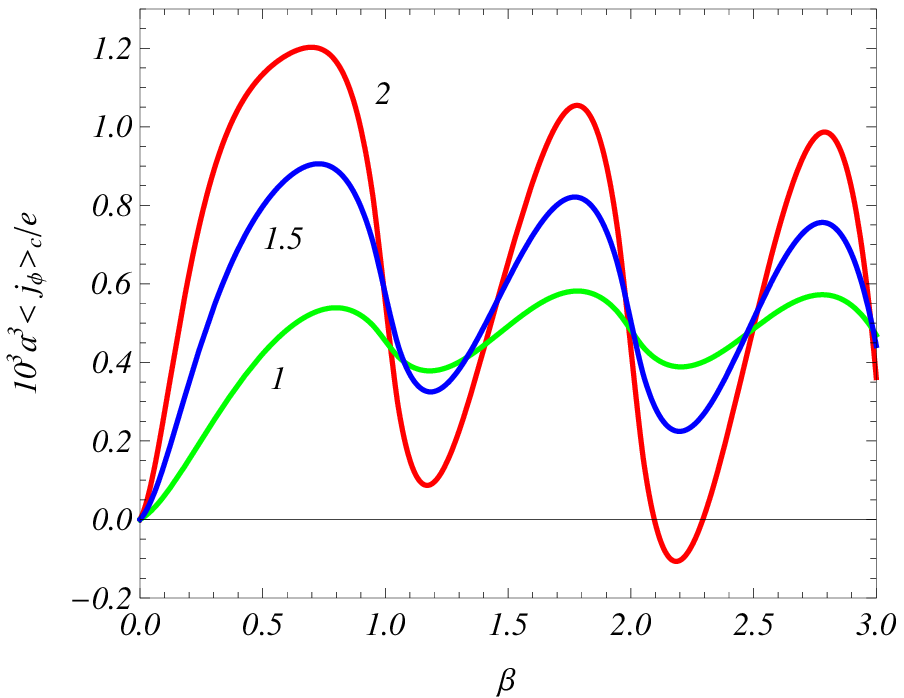,width=7.cm,height=5.5cm} & \quad %
\epsfig{figure=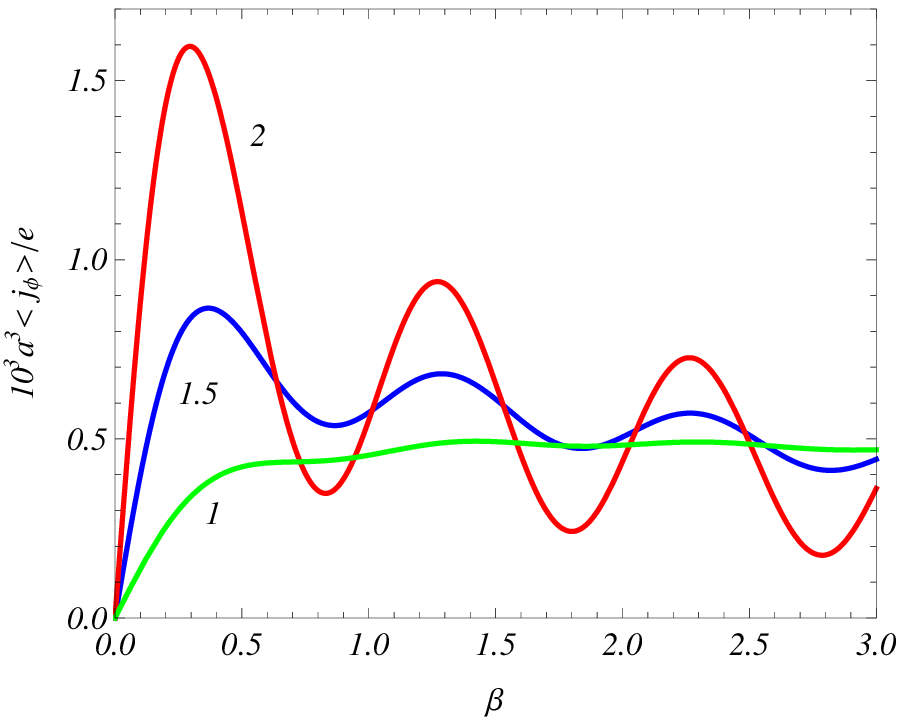,width=7.cm,height=5.5cm}%
\end{tabular}%
\end{center}
\caption{The core-induced contribution (left panel) and the total VEV (right
panel) of the current density for a $D=3$ minimally coupled massless scalar
field in the flower-pot model with an interior homogeneous magnetic field,
as a function of the parameter $\protect\beta $. The numbers near the curves
are the corresponding values of $q$ and the graphs are plotted for $r/a=2$.}
\label{fig5}
\end{figure}

In figure \ref{fig5}, the core-induced part (left panel) and the total VEV
of the current density (right panel) are depicted as functions of $\beta $
for $D=3$ minimally coupled massless scalar field. The graphs are plotted
for fixed value $r/a=2$ in the flower-pot model with the vector potential
distribution given by Eq. (\ref{A2homo}) and the numbers near the curves are
the corresponding values of the parameter $q$.

For the model (\ref{A2homo}) and for $D=3$, we have also numerically
investigated the behavior of the current density as a function of the radial
distance for $q=1.5$ and for two values of the magnetic flux corresponding
to $\beta =0.25$ and $\beta =1.75$. For both these values, the ratio $%
\left\langle j_{\phi }\right\rangle _{c}/e$ is positive and monotonically
decreasing with increasing $r$. It diverges on the boundary inversely
proportional to the distance from the boundary, which is in agreement with
the general analysis of the previous section. The zero-thickness cosmic
string part, $\left\langle j_{\phi }\right\rangle _{0}/e$, is positive for $%
\beta =0.25$ and negative for $\beta =1.75$, being monotonic in both cases.
As a result, in the case $\beta =0.25$ the ratio $\left\langle j_{\phi
}\right\rangle /e$ for the total current is positive monotonically
decreasing function of $r$. For $\beta =1.75$ the total VEV is positive near
the boundary and negative at large distances. It vanishes at $r/a\approx
3.15 $ and takes its minimum value at $r/a\approx 4$.

The third exactly solvable model corresponds to the vector potential%
\begin{equation}
A_{2}(r)=-qA_{\phi }r_{M},\;r<a,  \label{A2colu}
\end{equation}%
where $A_{\phi }=\mathrm{const}$ is the corresponding physical component.
For $D=3$ model and in the interior Minkowski coordinates $(t,r_{M},\phi
_{M},z)$, the nonzero component of the magnetic field is given by $%
B_{z}=A_{\phi }/r_{M}$. A charged particle with magnetic moment in this type
of magnetic field has been considered in Ref. \cite{Bord93}. From the
continuity condition for the vector potential at the core boundary, for the
covariant component in the exterior region, we get $A_{2}=-aA_{\phi }$. In
terms of the parameter $\beta $ one has $eA_{\phi }=-q\beta /a$. With the
function (\ref{A2colu}), the regular solution of the radial equation (\ref%
{radeq}) is expressed as%
\begin{equation}
R_{n}(r,\gamma )=C_{2}r_{M}^{|n|}e^{-hr_{M}/a}M(|n|+nq\beta
/h+1/2,2|n|+1,2hr_{M}/a),  \label{Rncolu}
\end{equation}%
where $h=(q^{2}\beta ^{2}-a^{2}\gamma ^{2})^{1/2}$ and the coefficient $C_{2}
$ is defined from Eq. (\ref{NormR}). For the function in the expression (\ref%
{j2c}) of the core-induced part of the current density we get
\begin{eqnarray}
p_{n}(ix) &=&q|n|+2\xi (q-1)-q\lambda (x)+q\left[ \lambda (x)+\frac{n\beta }{%
|n|+1/2}\right]   \notag \\
&&\times \frac{M(|n|+n\beta /\lambda (x)+3/2,2|n|+2,2\lambda (x))}{%
M(|n|+n\beta /\lambda (x)+1/2,2|n|+1,2\lambda (x))},  \label{pixcolu}
\end{eqnarray}%
with the notation%
\begin{equation}
\lambda (x)=\sqrt{(ax/q)^{2}+\beta ^{2}}.  \label{lamb}
\end{equation}%
Similarly to the previous case, for large magnetic fluxes, $|n_{0}|\gg
1/(r/a-1)$, the leading term in the expansion of Eq. (\ref{pixcolu})
vanishes and the next to the leading term does not depend on $n_{0}$. As a
result, to the leading order, the current density is a periodic function of
the magnetic flux.

\begin{figure}[tbph]
\begin{center}
\begin{tabular}{cc}
\epsfig{figure=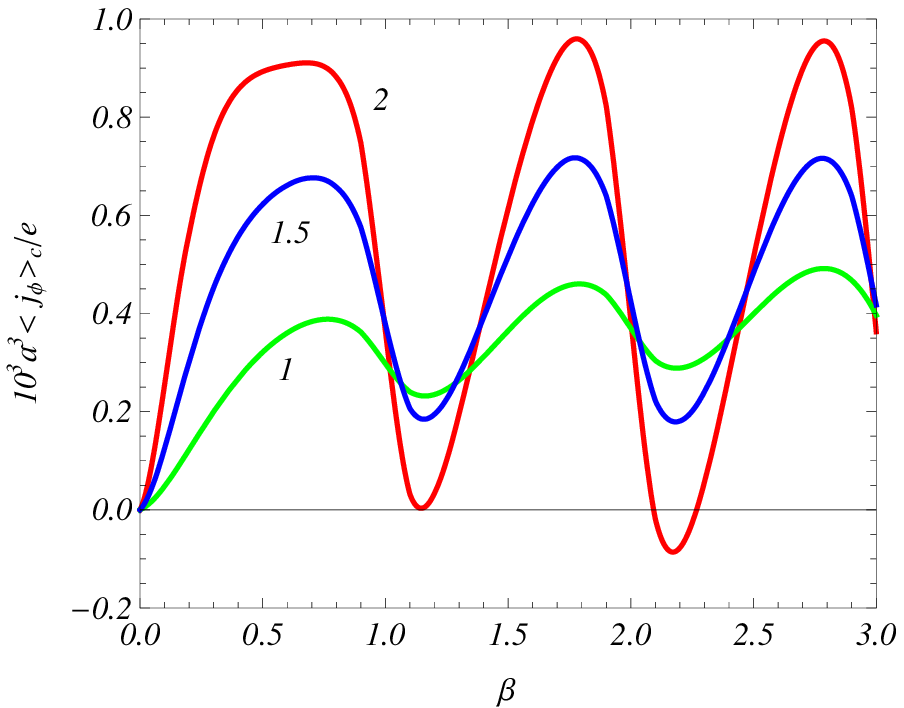,width=7.cm,height=5.5cm} & \quad %
\epsfig{figure=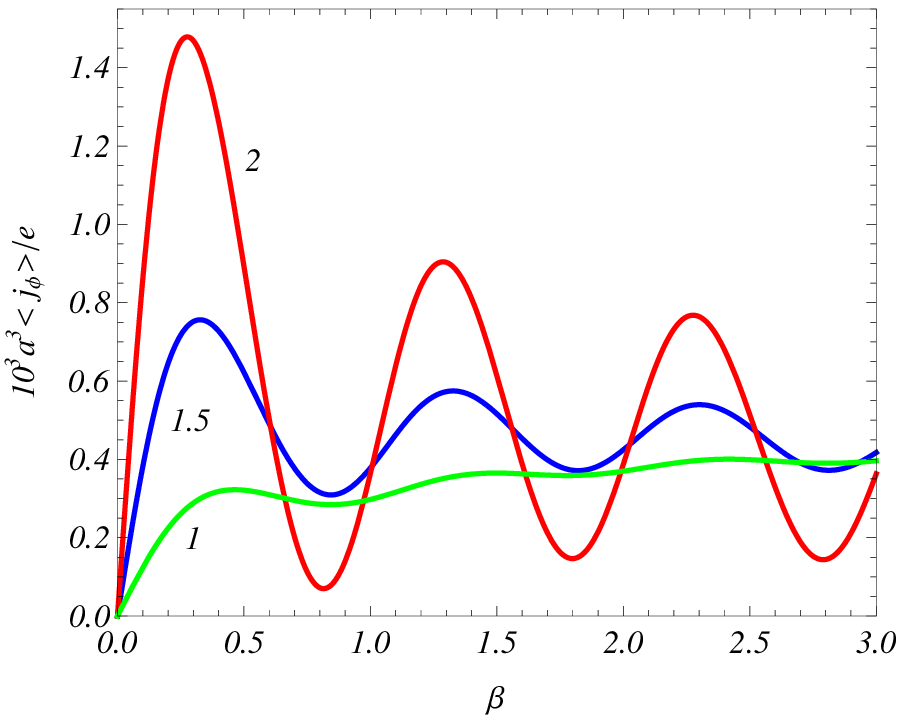,width=7.cm,height=5.5cm}%
\end{tabular}%
\end{center}
\caption{The same as in figure \protect\ref{fig5} for the flower-pot model
with the gauge field distribution given by Eq. (\protect\ref{A2colu}).}
\label{fig6}
\end{figure}

In figure \ref{fig6}, the core-induced part (left panel) and the total VEV
(right panel) of the current density are depicted as functions of $\beta $
for $D=3$ minimally coupled massless scalar field. The graphs are plotted
for fixed value $r/a=2$ in the flower-pot model with the gauge field
distribution given by Eq. (\ref{A2colu}). As before, the numbers near the
curves are the corresponding values of the parameter $q$.

\subsection{Ballpoint pen model}

For this model the interior geometry is described by a positive constant
curvature space. In this case we have%
\begin{equation}
u(r)=qb\sin (r_{i}/b),  \label{uBP}
\end{equation}%
where $r_{i}$ is the interior radial coordinate. From the continuity of the
metric tensor component $g_{11}$ we find $r_{i}=r-r_{c}$. From the condition
(\ref{ua}) one gets two possible values of the constant $r_{c}=r_{cj}$, $%
j=1,2$, which are%
\begin{eqnarray}
r_{c1} &=&a-b\arcsin [a/(qb)],  \notag \\
r_{c2} &=&a-b\left[ \pi -\arcsin (a/(qb))\right] .  \label{r12BP}
\end{eqnarray}%
For the corresponding surface energy-momentum tensor we find%
\begin{equation}
8\pi G\tau _{i}^{k}=\frac{\pm \sqrt{q^{2}-a^{2}/b^{2}}-1}{a}\delta
_{i}^{k},\;i,k=0,3,\ldots ,D,  \label{SurfBP}
\end{equation}%
where the upper and lower signs correspond to $r_{c1}$ and $r_{c2}$,
respectively. For the upper sign and in the special case%
\begin{equation}
a/b=\sqrt{q^{2}-1},  \label{Sp}
\end{equation}%
the surface energy-momentum tensor vanishes. In this case the first
derivatives of the metric tensor are continuous as well. The corresponding
model has been discussed in Ref. \cite{Hisk85} (see also Refs. \cite%
{Alle90,Khus99}). Note that, in the case $D=2$, the geometry under
consideration models a graphitic cone with a spherical cup, long wavelength
properties of which are well described by a $(2+1)$-dimensional
field-theoretical model. The corresponding opening angle is given by $\phi
_{0}=2\pi (1-N_{c}/6)$, where $N_{c}=1,2,\ldots ,5$ is the number of sectors
removed from the planar graphene sheet. All these angles have been
experimentally observed \cite{Kris97}.

In order to have an exactly solvable problem, for the gauge field
configuration in the interior region we shall take the model given by Eq. (%
\ref{A2shell}). The regular solution of the radial equation (\ref{radeq}) in
the interior region is expressed in terms of the associated Legendre
function
\begin{equation}
R_{n}(r,\gamma )=CP_{\nu (\gamma )-1/2}^{|n|}(\cos (r_{i}/b)),  \label{RnBP}
\end{equation}%
where%
\begin{equation}
\nu (\gamma )=\sqrt{b^{2}\gamma ^{2}-2\xi +1/4}.  \label{nu}
\end{equation}%
For the function $p_{n}(ix)$ in (\ref{j2c}) this gives%
\begin{equation}
p_{n}(ix)=2\xi (\pm \sqrt{q^{2}-a^{2}/b^{2}}-1)-\frac{a^{2}}{qb^{2}}\frac{%
P_{\nu (ix)-1/2}^{|n|\prime }(\pm \sqrt{1-a^{2}/q^{2}b^{2}})}{P_{\nu
(ix)-1/2}^{|n|}(\pm \sqrt{1-a^{2}/q^{2}b^{2}})},  \label{pixBP}
\end{equation}%
where the upper and lower signs correspond to the models with $r_{c}=r_{c1}$
and $r_{c}=r_{c2}$, respectively. If the derivative of the metric tensor is
continuous as well, then the expression is simplified to%
\begin{equation}
p_{n}(ix)=-\frac{q^{2}-1}{q}\frac{P_{\nu (ix)-1/2}^{|n|\prime }(1/q)}{P_{\nu
(ix)-1/2}^{|n|}(1/q)}.  \label{pixBPSp}
\end{equation}%
Note that, though the function $\nu (ix)$ can be either real or purely
imaginary, the function $P_{\nu (ix)-1/2}^{|n|}(1/q)$ is always real because
of the property $P_{\nu (ix)-1/2}^{|n|}(1/q)=P_{-\nu (ix)-1/2}^{|n|}(1/q)$
for the associated Legendre function.

\section{Conclusion}

\label{sec:Conc}

In the present paper we have investigated the finite core effects on the
vacuum current, induced by magnetic fluxes, for a massive scalar field in
the geometry of a straight cosmic string. For the interior structure of the
core we have considered a general static cylindrically symmetric geometry
with an arbitrary distributed gauge field flux along the string axis. For
generality, we have also assumed the presence of the surface energy-momentum
tensor on the core boundary. In the region outside the core the geometry is
described by the standard conical line element with a planar angle deficit
and for the gauge field configuration we have taken the vector potential
given by Eq. (\ref{Ak}) with $A_{2}(r)=\mathrm{const}$. Though the
corresponding field strength vanishes, the magnetic field inside the core
induces a nonzero VEV for the azimuthal current in the exterior region. This
current is a consequence of two types of effects: the Aharonov-Bohm like
effect and the direct interaction of the quantum field with the magnetic
field inside the core. In models with an impenetrable core the first effect
is present only and the induced current is a periodic function of the
magnetic flux inside the core with the period equal to the quantum flux. For
penetrable cores, the both type of effects contribute and the current
density, in general, is not a periodic function of the magnetic flux.

In the region outside the core, all the properties of the quantum vacuum can
be deduced from the corresponding two-point function. As such we have
considered the Hadamard function. In particular, the VEV of the current
density is obtained by using Eq. (\ref{jlVEV}). For the evaluation of the
Hadamard function we have used the direct summation over a complete set of
modes for a scalar field. In cylindrical coordinates and for scattering
states the radial parts in the mode functions are given by Eq. (\ref{fr}),
where the coefficients are determined from the matching conditions at the
core boundary. The normalization condition determines the value of the
interior radial function at the boundary (see Eq. (\ref{NormR})). In
addition to the scattering modes, there can be also bound states for which
the radial function at large distances from the core decreases
exponentially. The possible bound states are determined by Eq. (\ref{bseq})
and the normalization of the corresponding interior function is given by Eq.
(\ref{Normbs}). We have explicitly decomposed the Hadamard function into two
contributions. The first one, given by Eq. (\ref{G0}), corresponds to the
geometry of a zero thickness cosmic string and magnetic flux and the second
one is induced by the nontrivial core structure. For the first contribution,
a closed expression (\ref{G0c}) is provided which is convenient for the
investigation of the expectation values of various local characteristics of
the vacuum state (field squared, energy-momentum tensor, current density).
This expression generalizes various special cases previously discussed in
the literature. The core-induced effects are encoded in the part of the
Hadamard function given by Eq. (\ref{Gc}). It contains both the
contributions from the scattering and bound states. The specific properties
of the core appear through the function $p_{n}(iz)$.

The VEV of the current density is decomposed into the zero-thickness cosmic
string part, Eq. (\ref{j20b}), and the contribution coming from the finite
core, given by Eq. (\ref{j2c}). The only nonzero component corresponds to
the azimuthal current. The zero-thickness part in the current density is a
periodic function of the magnetic flux inside the core with the period equal
to the quantum flux. For penetrable cores, the core-induced contribution, in
general, is not periodic. The physical reason for this is the direct
interaction of the quantum field with the magnetic field inside the
penetrable core. For a massive field, at large distances from the string, $%
mr\gg 1$, both parts in the current density are exponentially small. For $q>2
$ the idealized string part dominates at large distances and the relative
contribution of the finite core effects are suppressed by the factor $%
e^{-2mr[1-\sin (\pi /q)]}/(mr)^{3/2}$. In the case $q\leqslant 2$ and for a
massive field, the contribution of the finite core effects is of the same
order as that coming from $\left\langle j_{\phi }\right\rangle _{0}$ if $%
am\sim 1$. For a massless field, the part corresponding to the geometry of
the zero thickness cosmic string behaves as $1/r^{D}$ and the relative
contribution of the finite core effects at large distances is suppressed by
the factor $(a/r)^{2\sigma _{f}}$, where $\sigma _{f}$ is defined by Eq. (%
\ref{sigma}). On the core boundary, the VEV of the current density, in
general, diverges. In order to find the leading term in the asymptotic
expansion of the current density over the distance from the core boundary,
we have provided the leading terms in the uniform asymptotic expansion of
the function $p_{n}(ix)$, for large values of $|n|$. The coefficients in the
asymptotic expansion are found by using Eq. (\ref{Eqyn}), which is obtained
from the equation for the interior radial function. If the vector potential
has discontinuity at the boundary, then the leading term in the expansion of
the current density near the boundary is given by Eq. (\ref{j2cnear}) and
the current density diverges as $1/(r-a)^{D-1}$. If the vector potential is
continuous at the boundary, the leading term vanishes and the asymptotic
expansion starts with the term of the order $1/(r-a)^{D-2}$. In special
cases of the interior geometry, the coefficient of the latter may vanish as
well. The part $\left\langle j_{\phi }\right\rangle _{0}$ is finite on the
boundary and, hence, near the boundary the current density, in general, is
dominated by the core-induced contribution. In models where the function $%
p_{n}(ix)$ does not depend on $\beta $, for large values of the magnetic
flux, the core-induced contribution in the current density, to the leading
order, coincides with the corresponding result for an impenetrable
cylindrical shell with Dirichlet boundary condition and the leading term is
periodic with the period equal to the quantum flux. The VEV of the total
current density, in general, is not a monotonic function of the distance
from the core boundary. By using the Maxwell semiclassical equation, we have
also investigated the magnetic field generated by the vacuum currents. In
particular, the core-induced contribution in the field strength is given by
Eq. (\ref{F21c}).

As applications of the general results we have considered four examples of
the core structure and the distribution of the gauge field. In the first
example, the core is impenetrable for the quantum field under consideration
and is modeled by a cylindrical surface on which the field operator obeys
Robin boundary condition. The core-induced contribution to the vacuum
current density is obtained from the general expression (\ref{j2c}) with the
function $p_{n}(ix)=\sigma a$. In this case, the current density is a
periodic function of the magnetic flux with the period equal to the quantum
flux. For the second example, we have considered the interior geometry being
Minkowskian (flower-pot model). The corresponding surface energy-momentum
tensor is obtained from the matching condition at the core boundary and is
given by Eq. (\ref{FPemt}). For the gauge field distribution we have
discussed three exactly integrable models. For $D=3$ they correspond to the
Dirac delta function type magnetic field located on a cylindrical shell, to
a homogeneous magnetic field inside the core and to a magnetic field
proportional to $1/r$. The corresponding functions are given by Eqs. (\ref%
{pixshell}), (\ref{pnxhom}) and (\ref{pixcolu}), respectively. In these
examples, as a consequence of the direct interaction of the quantum field
with the interior magnetic field, the current density is not a periodic
function of the magnetic flux. However, for large fluxes, the leading terms
in the corresponding asymptotic expansions are periodic with period equal to
the quantum flux. As a model for the interior geometry we have also
considered the ballpoint pen model with a constant curvature space. In this
case two possible matching conditions are obtained with the surface
energy-momentum tensors given by Eq. (\ref{SurfBP}). In the special case (%
\ref{Sp}) the derivatives of the metric tensor are continuous at the core
boundary and the surface energy-momentum tensor vanishes. In the ballpoint
pen model, in order to have an exactly solvable problem, we have considered
the Dirac-delta-type distribution of the magnetic field on the core
boundary. The corresponding function in the expression for the core-induced
contribution to the current density is given by Eq. (\ref{pixBP}).

\section*{Acknowledgments}

The authors thank Conselho Nacional de Desenvolvimento Cient\'{\i}fico e
Tecnol\'{o}gico (CNPq) for the financial support. A. A. S. was supported by
the State Committee of Science of the Ministry of Education and Science RA,
within the frame of Grant No. SCS 13-1C040.

\appendix

\section{Integral representation}

\label{sec:Appendix}

In this section we derive an integral representation for the function $%
\mathcal{I}_{q}(\beta ,\Delta \phi ,z)$ defined by Eq. (\ref{Ical}). By
taking into account Eq. (\ref{betdec}), we obtain%
\begin{equation}
\mathcal{I}_{q}(\beta ,\Delta \phi ,x)=e^{-iqn_{0}\Delta \phi }\mathcal{I}%
_{q}(\beta _{f},\Delta \phi ,x).  \label{Icalrel}
\end{equation}%
From here it follows that, without loss of generality, in the evaluation
below we can assume that $0\leqslant \beta <1$. We use the integral
representation for the modified Bessel function \cite{Abra72}:%
\begin{equation}
I_{\beta _{n}}(z)=\frac{1}{\pi }\int_{0}^{\pi }dy\;\cos (\beta
_{n}y)e^{z\cos y}-\frac{\sin (\pi \beta _{n})}{\pi }\int_{0}^{\infty
}dye^{-z\cosh y-\beta _{n}y}.  \label{IInt}
\end{equation}%
Substituting this into the right-hand side of Eq. (\ref{Ical}), for the part
with the first integral in Eq. (\ref{IInt}) we use the formula%
\begin{equation}
\sum_{n=-\infty }^{+\infty }e^{ibn}=2\pi \sum_{n=-\infty }^{+\infty }\delta
(b-2\pi n).  \label{SumForm}
\end{equation}%
This gives%
\begin{equation}
\frac{1}{\pi }\int_{0}^{\pi }dy\,e^{z\cos y}\sum_{n=-\infty }^{+\infty
}e^{iqn\Delta \phi }\cos (\beta _{n}y)=\frac{1}{q}\sum_{n}e^{z\cos \left(
2\pi n/q-\Delta \phi \right) }e^{iq\beta \left( 2\pi n/q-\Delta \phi \right)
},  \label{1term}
\end{equation}%
where in the right-hand side the summation goes under the condition%
\begin{equation}
-q/2+\Delta \phi /\phi _{0}\leqslant n\leqslant q/2+\Delta \phi /\phi _{0}.
\label{sumcond}
\end{equation}%
If $-q/2+\Delta \phi /\phi _{0}$ or $q/2+\Delta \phi /\phi _{0}$ are
integers then the corresponding terms in the right-hand side of Eq. (\ref%
{1term}) should be taken with the coefficient 1/2.

In the part with the second integral in the right-hand side of Eq. (\ref%
{IInt}) we use the formula%
\begin{equation}
\sum_{n=-\infty }^{+\infty }e^{iqn\Delta \phi -\beta _{n}y}\sin (\pi \beta
_{n})=\frac{1}{2i}\sum_{j=+,-}je^{j\pi iq\beta }\frac{\cosh [qy(1-\beta
)]-\cosh (q\beta y)e^{-iq\left( \Delta \phi +j\pi \right) }}{\cosh (qy)-\cos
(q\left( \Delta \phi +j\pi \right) )}.  \label{2term}
\end{equation}%
Combining Eqs. (\ref{Icalrel}), (\ref{1term}) and (\ref{2term}), for the
general case of $\beta $ we obtain the following integral representation
\begin{eqnarray}
\mathcal{I}_{q}(\beta ,\Delta \phi ,z) &=&\frac{1}{q}\sum_{n}e^{z\cos \left(
2\pi n/q-\Delta \phi \right) }e^{i\beta (2\pi n-q\Delta \phi )}-\frac{%
e^{-iqn_{0}\Delta \phi }}{2\pi i}\sum_{j=+,-}je^{ji\pi q\beta _{f}}  \notag
\\
&&\times \int_{0}^{\infty }dy\,\frac{\cosh [qy(1-\beta _{f})]-\cosh (q\beta
_{f}y)e^{-iq\left( \Delta \phi +j\pi \right) }}{e^{z\cosh y}[\cosh (qy)-\cos
(q\left( \Delta \phi +j\pi \right) )]}.  \label{Ical1}
\end{eqnarray}%
For the special case $\beta =0$ this formula reduced to the one given in
Ref. \cite{Beze12}.

Taking $\Delta \phi =0$, from (\ref{Ical1}) one gets%
\begin{eqnarray}
\sum_{n=-\infty }^{+\infty }I_{\beta _{n}}(z) &=&\frac{2}{q}%
\sideset{}{'}{\sum}_{n=0}^{[q/2]}\cos (2\pi n\beta )e^{z\cos \left( 2\pi
n/q\right) }-\frac{1}{\pi }\int_{0}^{\infty }dy  \notag \\
&&\times \frac{\sin (\pi q\beta _{f})\cosh [qy(1-\beta _{f})]+\sin (\pi
q(1-\beta _{f}))\cosh (q\beta _{f}y)}{e^{z\cosh y}[\cosh (qy)-\cos (\pi q)]}.
\label{Isum}
\end{eqnarray}%
where the prime on the sign of the sum means that the terms $n=0$ and $n=q/2$
(if $q$ is an even integer) should be taken with the coefficient 1/2.

A similar formula for the series in Eq. (\ref{j20}) can be obtained by using
Eq. (\ref{Isum}) and the relation%
\begin{equation}
q(n+\beta )I_{\beta _{n}}(z)=-z\partial _{z}I_{\beta _{n}}(z)+zI_{q|n+\beta
-1/q|}(z),  \label{RelI}
\end{equation}%
valid for $n\neq 0$ and $0<\beta <1$. For $\beta >1/q$ this gives%
\begin{equation}
q\sum_{n=-\infty }^{+\infty }(n+\beta )I_{\beta _{n}}(z)=-z\partial
_{z}\sum_{n=-\infty }^{+\infty }I_{\beta _{n}}(z)+z\sum_{n=-\infty
}^{+\infty }I_{q|n+\beta -1/q|}(z).  \label{RelSer}
\end{equation}%
Applying the formula (\ref{Isum}) (with the replacement $\beta \rightarrow
\beta -1/q$ for the last series in (\ref{RelSer})), after some
transformations we get%
\begin{eqnarray}
\sum_{n=-\infty }^{+\infty }(n+\beta )I_{\beta _{n}}(z) &=&\frac{2z}{q^{2}}%
\sum_{n=1}^{[q/2]}\sin \left( 2\pi n/q\right) \sin (2\pi n\beta )e^{z\cos
\left( 2\pi n/q\right) }  \notag \\
&&+\frac{z}{\pi q}\int_{0}^{\infty }dy\,\frac{\sinh ye^{-z\cosh y}f(q,\beta
_{f},y)}{\cosh (qy)-\cos (\pi q)},  \label{Isum2}
\end{eqnarray}%
with the function $f(q,\beta _{f},y)$ defined by Eq. (\ref{fq}). Note that
for $q$ being an even integer the contribution of the term $n=q/2$, in Eq. (%
\ref{Isum2}), vanishes.

For $\beta <1/q$ one has the relation%
\begin{equation}
q\sum_{n=-\infty }^{+\infty }(n+\beta )I_{\beta _{n}}(z)=\frac{2}{\pi }\sin
[\pi (1-q\beta )]zK_{1-q\beta }(z)-z\partial _{z}\sum_{n=-\infty }^{+\infty
}I_{\beta _{n}}(z)+z\sum_{n=-\infty }^{+\infty }I_{q|n+\beta -1/q+1|}(z).
\label{RelI2}
\end{equation}%
Again, by using Eq. (\ref{Isum}) and the integral representation%
\begin{equation}
K_{1-q\beta }(z)=\int_{0}^{\infty }dye^{-z\cosh y}\cosh [(1-q\beta )y],
\label{Kint}
\end{equation}%
we obtain the same representation (\ref{Isum2}). Hence, the formula (\ref%
{Isum2}), with $\beta _{f}$ defined by Eq. (\ref{betdec}), is valid for
general values of $\beta $.

\end{document}